\documentclass{IEEEtran}
\usepackage{cite}
\usepackage{amsmath,amssymb,amsfonts}
\usepackage{algorithmic}
\usepackage{graphicx}
\usepackage{textcomp}
\usepackage{xcolor}

\newcommand{\dbar}[1]{\overline{\overline{#1}}}

\def\BibTeX{{\rm B\kern-.05em{\sc i\kern-.025em b}\kern-.08em
    T\kern-.1667em\lower.7ex\hbox{E}\kern-.125emX}}
\begin{document}
\title{All-Angle Nonlocal Metasurfaces on Demand: Universal Realization of Normal Susceptibilities via Multilayered Printed-Circuit-Board (PCB) Cascades}
\author{Amit Shaham, \IEEEmembership{Graduate Student Member, IEEE,} and Ariel Epstein, \IEEEmembership{Senior Member, IEEE}
\thanks{Manuscript submitted on December 22, 2024.}
\thanks{The authors are with the Andrew and Erna Viterbi Faculty of Electrical and Computer Engineering, Technion--Israel Institute of Technology, Haifa, 3200003, Israel (email: samitsh@campus.technion.ac.il; epsteina@ee.technion.ac.il).}}

\maketitle

\begin{abstract}
Embedding normal susceptibilities in metasurfaces (MS), in tandem with their tangential counterparts, greatly enriches their spatial dispersion. Particularly, judicious siphoning of the microscopic nonlocality associated with such enhanced meta-atoms facilitates global control over the MS response across the entire angular range, specifically at challenging near-grazing scenarios. In this paper, we introduce a rigorous closed-form methodology to realize such intricate mixtures of tangential and normal components via a highly practical platform---printed-circuit-board- (PCB) compatible cascaded admittance sheets. To this end, we derive a universal all-angular link between this MS-level composite, which leverages macroscopic nonlocality of multiple reflections, to its underlying meta-atom-level susceptibilities. We demonstrate this scheme by devising a PCB all-angle-transparent generalized Huygens' MS radome and an all-angle perfect-magnetic-conductor (PMC) PCB MS. Validated in simulation and experiment, our results pave the path toward a new insightful paradigm for studying and engineering nonlocal metadevices, e.g., optical analog computers and spaceplates.
\end{abstract}

\begin{IEEEkeywords}
Artificial magnetic conductors (AMC), bianisotropy, generalized Huygens' condition (GHC), high-impedance surfaces (HIS), metasurfaces (MS), nonlocality, radomes, spatial dispersion, spatial filtering.
\end{IEEEkeywords}

\section{Introduction}
\label{Sec:Introduction}
\IEEEPARstart{M}{odern} metasurfaces (MS) have been celebrating more than a decade of successful applicability and performance, from microwave frequencies to optical wavelengths \cite{Glybovski2016}. Thus far, these thin dense arrangements of miniature dielectric and metallic inclusions have demonstrated efficient and versatile wavefront manipulation, including beamforming, antenna enhancement, and polarization control.

One prominent class of MSs, termed Huygens’ MSs (HMS) \cite{Pfeiffer2013,Monticone2013,Pfeiffer2013Cascaded,Selvanayagam2013,Pfeiffer2013Millimeter,PfeifferNano2014,Wong2014,Epstein2016}, is devised by carefully positioning Huygens’ scatterers to fulfill a predetermined local field transformation with minimal reflection, for instance, anomalous refraction. A Huygens’ scatterer consists of collocated tangential electric and magnetic responses, balanced such that reflection off the former destructively interferes with that off the latter \cite{Love1976,Jin2010,Decker2015}. From the \emph{meta-atom viewpoint}, this balance can be microscopically accomplished, e.g., by a set of subwavelength polarizable elements, for instance, loops and wires \cite{Pfeiffer2013,Selvanayagam2013,Wong2014}. However, for more convenient practical realizations, e.g., printed-circuit-board (PCB) compatible admittance sheet cascades at microwave frequencies \cite{Monticone2013,Pfeiffer2013Cascaded,Pfeiffer2013Millimeter}, a \emph{macroscopic MS approach} would be more apt: the balance requisite is effectively emulated by symmetric (electric) and antisymmetric (magnetic) modes supported by the structure \cite{Pfeiffer2013Millimeter,Epstein2016OBMS}, as also applicable in optics \cite{Decker2015}.

Powerful as it is, Huygens’ condition (HC) in its \emph{tangentially polarizable} form \cite{Pfeiffer2013,Monticone2013,Pfeiffer2013Cascaded,Selvanayagam2013,Pfeiffer2013Millimeter,PfeifferNano2014,Wong2014,Epstein2016} is fundamentally limited as far as planar wide-angle reflectionless utilities, such as antireflective coatings \cite{Dobrowolski2002,Chattopadhyay2010,He2018}, antenna radomes \cite{Finley1956,Munk1971,Pelton1974,He2020,Goshen2024}, and other perfectly matched devices \cite{Radi2015,Gok2016,RuizGarcia2024}, are concerned. The reason is that HC is tailored at a single angle of incidence (typically normal), such that deviation from this angle inevitably alters the ratio between the transverse electric and magnetic fields (namely, the wave impedance), and, in turn, leads to unbalanced ratio of induced dipoles that inflicts undesired reflection \cite{Holloway2005,Epstein2016}.

More generally, angular sensitivity of MS scattering and absorption intrinsically appears beyond HMS functionalities. It originates from \emph{nonlocality} \cite{Asadchy2017PhD,Overvig2022,Shastri2023}, an elementary property of waves where the host medium responds not only to the local fields at a certain point in space, but also to the fields in an extended region around it. Consequently, the propagation and scattering characteristics of nonlocal media are \emph{angle} (or \emph{wavevector}) \emph{dependent}, i.e., \emph{spatially dispersive}.

Similarly to the case of HMSs, spatial dispersion is considered an irritant for other purposes, such as artificial perfect magnetic conductors (PMC, or high-impedance surfaces, HIS) \cite{Sievenpiper1999,Simovski2004,Feresidis2005,Monorchio2006,Hashemi2013}
and wide-angle impedance matching (WAIM) of antenna arrays \cite{Magill1966,Cameron2015,Soltani2024}, as it strands their operation to a narrow band of angles. On the other hand, rather than expelling it, emergent trends in metamaterial and MS communities have exploited it to introduce novel functionalities, e.g., optical analog computers \cite{Silva2014,Zhu2017,Kwon2018,Momeni2021,Cordaro2023} and space-squeezing plates \cite{Guo2020,Chen2021,Reshef2021,Page2022,Shastri2022,Fernandez2024}; additional classical devices, such as angular filters \cite{Munk1971,Mailloux1976,Franchi1983,Ortiz2013,Shaham2022,Abdipur2024}, and other advanced wave-manipulating MSs, e.g., \cite{Pfeiffer2016,Epstein2016PRL,Xu2024}, utilize nonlocality as well (even if tacitly, at times).

In light of the above, it is clear that careful and systematic regulation of spatial dispersion should be generally practiced as a key design aspect in metamaterial and MS applications. Since naturally occurring materials are mostly limited to weak microscopic nonlocal effects \cite{Shastri2023}, such phenomena are often realized macroscopically, by means of artificial composites that leverage fundamental wave mechanisms, such as refraction and interference of multiple delayed reflections in multilayered media \cite{Gerken2003,Silva2014,Ishimaru2017,Chen2021,Shastri2022,Abdipur2024}; coupling to guided, surface, and leaky waves supported by interconnected, strongly intercoupled, or spatially modulated regions \cite{Ortiz2013,Monticone2015,Kwon2018}; and their combinations \cite{Overvig2022,Xu2024}.

Diverse as these physical principles may seem, the prevalent procedures to engineer them are quite similar. First, a goal functionality (angular or wavevector response) is formulated in terms relevant to the physical phenomenon under consideration (e.g., scattering coefficients, permittivity and permeability, or surface impedance). Next, based on the qualitative features demanded by the target operation (for instance, symmetry), a generic meta-atom or unit-cell geometry is proposed. Thereafter, its dimensions are optimized, typically through parametric sweeps in numerical or full-wave solvers, to meet these specifications to the best degree possible.

Unfortunately, while such design techniques have indeed facilitated the pioneering demonstrations reported above (and more), they naturally suffer several fundamental limitations. In the vast majority of cases, the actual response strongly deviates from the intended one outside a small range of angles around the optical axis. In other words, their acceptance angles or numerical apertures (NA) are typically modest (a few tens of degrees at most). The reason is that the chosen meta-atom geometry is not perfectly compatible with the desired electromagnetic response, in the sense that it cannot be tuned to precisely maintain the strength or balance of electromagnetic reactions across the entire angular range (\emph{a fortiori} when paraxial approximations are involved). Furthermore, this implementation methodology is not modular, i.e., the meta-atom configuration is chosen individually for a specific functionality, not for a set of possible functionalities; this burdens the design process and hinders possible reconfigurable extensions.

In search of a remedy for a tightly related issue, we have recently derived the generalized Huygens' condition (GHC) \cite{Shaham2023}, a simple yet powerful generalization of the standard HC \cite{Pfeiffer2013,Monticone2013,Pfeiffer2013Cascaded,Selvanayagam2013,Pfeiffer2013Millimeter,PfeifferNano2014,Wong2014,Epstein2016} that suppresses reflection for \emph{all} angles of incidence. Specifically, at the \emph{meta-atom level}, we have found that the aforementioned \emph{tangential} balance between the electric and magnetic responses (HC, i.e., zero reflection, at normal incidence) would not suffice by itself: another distinct balance between tangential and \emph{normal} components must be enforced to overturn reflection at \emph{grazing} incidence as well (another particular HC), to thus ensure omnidirectional performance. 

In fact, introducing such normal susceptibilities in the GHC provides enhanced means to channel the intrinsic nonlocality imbued in Maxwell's equations (spatial derivative relations between field components) \cite{Shaham2023}, as also generally observed in other past studies \cite{Holloway2005,Holloway2009,Hashemi2013,Ortiz2013,Cameron2015,Zaluski2016,Pfeiffer2016,Albooyeh2017,Achouri2018,Achouri2020,Momeni2021,delRisco2021,Shaham2021,Shaham2022,Smy2022,Tiukuvaara2022}. So far, these unconventional components have mainly been implemented via strongly resonant subwavelength in-plane loops or short out-of-plane wires, pillars, or slots. However, further exploring the grazing-angle Huygens' condition at the \emph{MS level} \cite{Shaham2023}, we have astonishingly found that it can be nonintuitively enforced in a simple PCB-compatible cascade of two admittance sheets separated by a standard dielectric substrate. That is, despite the absence of normally polarizable components in each of the individual layers, we have rigorously shown that the finitely thick structure indeed macroscopically enacts effective normal susceptibilities. This demonstration of equivalence, in fact, heavily implies that two apparently distinct macroscopic nonlocal mechanisms---weak nonlocality that allows electrically polarizable materials to manifest effective magnetic response (meta-atom-level) \cite{Asadchy2017PhD}, and multiple reflections in stratified media (MS level) \cite{Shastri2023}---can exhibit identical scattering behavior for all angles.

In this paper, we generalize this concept by revealing an explicit universal nonlocal link between trilayered MS cascades of dielectrically separated admittance sheets and their equivalent homogenized representation via local surface susceptibilities, corresponding to the standard generalized sheet transition conditions (GSTCs) \cite{Idemen1990,Tretyakov2003,Kuester2003,Achouri2018}. This convenient closed-form connection unravels the role of each admittance layer in determining each of the underlying (angularly independent) susceptibility values, and, in turn, the entire angular response of the MS. Specifically, we observe that normal susceptibilities establish an inextricable part of such multilayered PCB devices, albeit having been overlooked by their standard design practices, most popularly HMSs \cite{Monticone2013,Pfeiffer2013Cascaded,Selvanayagam2013,Pfeiffer2013Millimeter,PfeifferNano2014,Epstein2016}, omega-bianisotropic MSs (OBMS) \cite{Epstein2016OBMS,Epstein2016PRL}, and chiral-bianisotropic MSs \cite{Pfeiffer2014}.

By this, not only do we unearth the equivalence between two fundamentally different agents of nonlocality (Maxwell's equations at the meta-atom level versus refraction and multiple reflections at the MS level), which is describable by a global framework (Sec.\ \ref{Sec:Theory}); we also, in effect, devise a holistic and malleable \emph{all-angle} scheme to realize intricate combinations of tangential and normal susceptibilities via a highly practical MS platform: PCB cascades of tangentially polarizable admittance sheets. We exemplify the usefulness, simplicity, and accuracy of our theory by rigorously designing two key functionalities: a \emph{perfect} generalized Huygens' MS radome transparent at all angles \cite{ShahamEuCAP} (Sec.\ \ref{Subsec:Radome}) and an all-angle artificial PMC MS (Sec.\ \ref{Subsec:PMC}). We deduce and discuss relevant insights that appear in light of this new approach and provide full-wave and experimental validation. Exhibiting excellent performance and versatility, our findings herein are expected to prelude a new paradigm, where various nonlocal mechanisms across the electromagnetic spectrum are unified under a coherent formalism concomitant of a straightforward realistic implementation.

\begin{figure*}
\centerline{\includegraphics[width=\textwidth]{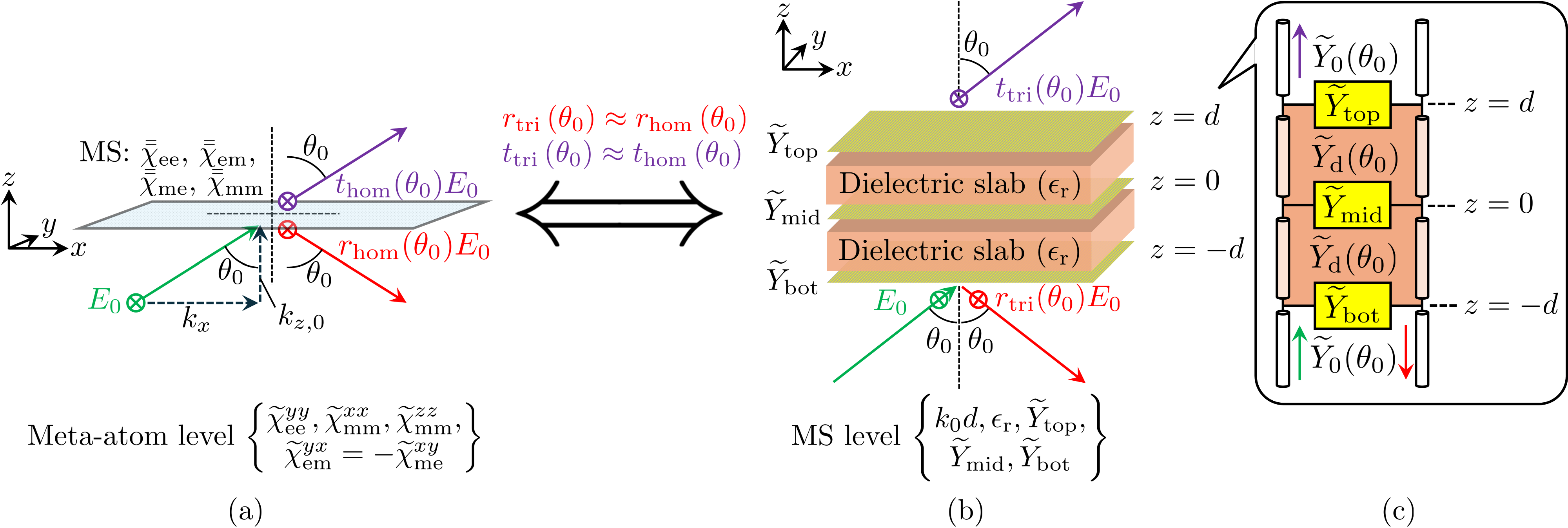}}
\caption{(a) Scattering configuration off a thin homogenized sheet ($z=0$) at the meta-atom level (Sec.\ \ref{Subsec:MALevel}): incident (green), specularly reflected (red), and directly transmitted (purple) $y$-polarized TE plane waves, as evaluated at $z=0$; wavevector decomposition into tangential, $k_{x}=k_0\sin\theta_0$, and normal, $k_{z,0}=k_0\cos\theta_0$, components related to the angle of incidence $\theta_0$; surface susceptibility tensors ($\bar{\bar{\chi}}$) defined in (\ref{Eq:Suscept}) and (\ref{Eq:SusceptComponents}). (b) Scattering configuration off a trilayered MS-level PCB-compatible cascade of admittance sheets ($\widetilde{Y}_{\mathrm{bot}}$, $\widetilde{Y}_{\mathrm{mid}}$, and $\widetilde{Y}_{\mathrm{top}}$) separated by two thin dielectric substrates of relative permittivity $\epsilon_{\mathrm{r}}$ and thickness $d$ each (Sec.\ \ref{Subsec:MSLevel}): an incident $y$-polarized plane wave (green) is scattered into specular reflection (red) and direct transmission (purple). (c) Equivalent TL circuit model for the setup in (b): free-space and dielectric regions are modeled by corresponding (angularly dependent) TLs and admittance sheets serve as shunt loads. The \emph{all-angle} equivalence ``$\Leftrightarrow$'' between the meta-atom level in (a) and the MS level in (b) is derived in Sec.\ \ref{Subsec:Equivalence}.}
\label{Fig:Theoretical_Config}
\end{figure*}

\section{Theroy}
\label{Sec:Theory}

\subsection{Thin Homogenized Sheets (Meta-Atom Level)}
\label{Subsec:MALevel}
To construct the grounds on which we are to base our theory, we set out by prescribing the angular dependence of scattering off thin homogenized sheets at the \emph{meta-atom level}. Let us consider one such sheet located at the $z=0$ plane, as in Fig.\ \ref{Fig:Theoretical_Config}(a). Depending on the constituents incorporated in the MS, surface electric $\vec{P}_{\mathrm{s}}$ and magnetic $\vec{M}_{\mathrm{s}}$ polarizations are induced by the impinging fields. These polarizations, in turn, introduce discontinuities of the tangential fields across the MS plane, which are governed by the GSTCs \cite{Idemen1990,Tretyakov2003,Kuester2003,Achouri2018},
\begin{equation}
\label{Eq:GSTCs}
    \begin{aligned}
    \hat{z}\times\left( \vec{H}_{\mathrm{t}}^{+}-\vec{H}_{\mathrm{t}}^{-} \right)&=j\omega\vec{P}_{\mathrm{st}}-\hat{z}\times\vec{\nabla}_{\mathrm{t}}M_{\mathrm{s}z},\\
    \left( \vec{E}_{\mathrm{t}}^{+}-\vec{E}_{\mathrm{t}}^{-}\right)\times\hat{z}&=j\omega\mu_{0}\vec{M}_{\mathrm{st}}-\vec{\nabla}_{\mathrm{t}}\left( \frac{P_{\mathrm{s}z}}{\epsilon_{0}}\right)\times\hat{z}.
    \end{aligned}
\end{equation}
Here, $\hat{z}$ is the normal to the MS plane; ``$\mathrm{t}$'' subscript denotes vector components tangential to the MS plane; $\pm$ superscripts denote field values near the top, $z\to 0^{+}$, and bottom, $z\to 0^{-}$, facets, respectively; $\epsilon_0$ and $\mu_0$ are the permittivity and permeability of the free space surrounding the MS; and global harmonic time dependence of $e^{j\omega t}$ is assumed and suppressed.

For passive MSs, the polarizations, $\vec{P}_{\mathrm{s}}$ and $\vec{M}_{\mathrm{s}}$, are formed by homogenizing the dipole reaction of the MS elements to the external fields. We follow the standard assumption that this reaction is \emph{local}, namely, that each element reacts only to the total field \emph{in situ} (approximately), such that \cite{Idemen1990,Tretyakov2003,Kuester2003,Achouri2018}
\begin{equation}
\label{Eq:Suscept}
    \begin{aligned}
		\vec{P_{\mathrm{s}}}&=\epsilon_{0}\dbar{\chi}_{\mathrm{ee}}\cdot\vec{E}^{\mathrm{av}}+c^{-1}\dbar{\chi}_{\mathrm{em}}\cdot\vec{H}^{\mathrm{av}},\\	\vec{M}_{\mathrm{s}}&=\eta_{0}^{-1}\dbar{\chi}_{\mathrm{me}}\cdot\vec{E}^{\mathrm{av}}+\dbar{\chi}_{\mathrm{mm}}\cdot\vec{H}^{\mathrm{av}},
	\end{aligned}
\end{equation}
where $\dbar{\chi}_{\mathrm{ee}}$, $\dbar{\chi}_{\mathrm{mm}}$, $\dbar{\chi}_{\mathrm{em}}$, and $\dbar{\chi}_{\mathrm{me}}$ are the surface electric, magnetic, electromagnetic and magnetoelectric susceptibility tensors, respectively; $\vec{E}^{\mathrm{av}}=\frac{1}{2}\left(\vec{E}^{+}+\vec{E}^{-}\right)$ and $\vec{H}^{\mathrm{av}}=\frac{1}{2}\left(\vec{H}^{+}+\vec{H}^{-}\right)$ denote the average values of the total electric and magnetic fields at both sides of the MS, respectively; and $c=(\mu_0\epsilon_0)^{-1/2}$ and $\eta_0=(\mu_0/\epsilon_0)^{1/2}$ are the speed of light and wave impedance in free space, respectively.

In general, each entry or dyad of the $3\times 3$ susceptibility tensors in (\ref{Eq:Suscept}) corresponds to a Cartesian polarization component driven by a Cartesian field component. In practice, their values are directly determined by the materials and geometries of the meta-atoms constituting the MS, so that they serve as the primary degrees of freedom to control its operation. The susceptibilities can be spatially varied along the MS plane, i.e., $\dbar{\chi}\left(\vec{r}\right)$, as in \cite{Pfeiffer2013,Monticone2013,Pfeiffer2013Cascaded,Selvanayagam2013,Pfeiffer2013Millimeter,PfeifferNano2014,Wong2014,Epstein2016,Epstein2016OBMS,Epstein2016PRL}. Furthermore, the form of (\ref{Eq:Suscept}) may, in principle, be generalized into spatial convolution (or, equivalently, assume wavevector-dependent susceptibilities) to address inclusions that introduce stronger spatial dispersion \cite{Achouri2022,Overvig2022,Shastri2023,Rahmier2023}. For our needs herein, the \emph{spatially invariant} and \emph{local} form of (\ref{Eq:Suscept}) will suffice.

For simplicity, we focus on a specific $y$-polarized transverse-electric (TE, ${\partial_{y}\equiv 0}$ and ${E_{x}=E_{z}=H_{y}\equiv0}$) configuration featured by a homogeneous MS \cite{Achouri2018} of the form
\begin{equation}
\label{Eq:SusceptComponents}
    \begin{aligned}
    &\dbar{\chi}_{\mathrm{ee}}=
    \begin{bmatrix}
        0 &0  &0\\
        0   &\chi_{\mathrm{ee}}^{yy}    &0\\
        0   &0   &0
    \end{bmatrix}
    ,
    &\dbar{\chi}_{\mathrm{mm}}=
    \begin{bmatrix}
        \chi_{\mathrm{mm}}^{xx} &0  &0\\
        0   &0    &0\\
        0   &0   &\chi_{\mathrm{mm}}^{zz}
    \end{bmatrix},\\
    &\dbar{\chi}_{\mathrm{em}}=
    \begin{bmatrix}
        0 &0  &0\\
        \chi_{\mathrm{em}}^{yx}   &0    &0\\
        0   &0   &0
    \end{bmatrix}, &\dbar{\chi}_{\mathrm{me}}=
    \begin{bmatrix}
        0 &\chi_{\mathrm{me}}^{xy}  &0\\
        0   &0    &0\\
        0   &0   &0
    \end{bmatrix},
    \end{aligned}
\end{equation}
where $\chi_{\mathrm{ee}}^{yy}$, $\chi_{\mathrm{mm}}^{xx}$, $\chi_{\mathrm{mm}}^{zz}$, $\chi_{\mathrm{em}}^{yx}$, and $\chi_{\mathrm{me}}^{xy}$ are the tangential electric, tangential magnetic, normal magnetic, tangential electromagnetic (omega-bianisotropic), and tangential magnetoelectric (omega-bianisotropic) components \cite{Asadchy2018}. Moreover, to obtain reciprocal structures, which facilitate practical realization without external bias, we enforce reciprocity, i.e.,
\begin{equation}
 \label{Eq:Reciprocity}
    \dbar{\chi}_{\mathrm{ee}}^{\mathrm{T}}=\dbar{\chi}_{\mathrm{ee}},\; \dbar{\chi}_{\mathrm{mm}}^{\mathrm{T}}=\dbar{\chi}_{\mathrm{mm}},\; \dbar{\chi}_{\mathrm{me}}^{\mathrm{T}}=-\dbar{\chi}_{\mathrm{em}},
\end{equation}
where $\left(\cdot\right)^{\mathrm{T}}$ denotes matrix transposition \cite{Asadchy2018,Pozar2012,Achouri2018,Pfeiffer2016}. Particularly, (\ref{Eq:Reciprocity}) stipulates ${\chi_{\mathrm{em}}^{yx}=-\chi_{\mathrm{me}}^{xy}}$; hence, only $\chi_{\mathrm{em}}^{yx}$ will be mentioned throughout for brevity. We also consider lossless scenarios (low-loss in reality), for which \cite{Achouri2018}
\begin{equation}
\label{Eq:Lossless}
    \dbar{\chi}_{\mathrm{ee}}^{\mathrm{T}}=\dbar{\chi}_{\mathrm{ee}}^{*},\; \dbar{\chi}_{\mathrm{mm}}^{\mathrm{T}}=\dbar{\chi}_{\mathrm{mm}}^{*},\; \dbar{\chi}_{\mathrm{me}}^{\mathrm{T}}=\dbar{\chi}_{\mathrm{em}}^{*},
\end{equation}
where $\left(\cdot\right)^{*}$ denotes elementwise complex conjugation. Together, (\ref{Eq:Reciprocity}) and (\ref{Eq:Lossless}) state that the passive and lossless inclusions of interest to us possess purely real $\chi_{\mathrm{ee}}^{yy}$, $\chi_{\mathrm{mm}}^{xx}$, and $\chi_{\mathrm{mm}}^{zz}$ and purely imaginary $\chi_{\mathrm{em}}^{yx}$ values.

We illuminate the MS from below ($z<0$) by a $y$-polarized transverse electric (TE) plane wave [Fig.\ \ref{Fig:Theoretical_Config}(a)], described via
\begin{equation}
\label{Eq:EyInc}
    E_{y}^{\mathrm{inc}}\left(\vec{r}\right)=E_0e^{-j\left(k_{x}x+k_{z,0}z\right)},
\end{equation}
where $E_0$ is the amplitude at the origin, $k_{x}=k_0\sin\theta_0$ and $k_{z,0}=k_0\cos\theta_0$ are the tangential and normal wavenumbers in free space, $k_0=\omega/c$ is the wave number in free space, and $\theta_0$ is the angle of incidence. The 2D translationally invariant TE configuration (${E_{x}=E_{z}=H_{y}\equiv0}$ and ${\partial_{y}\equiv 0}$) provided by (\ref{Eq:SusceptComponents}) gives rise only to $y$-polarized specularly reflected and directly transmitted waves, i.e., no cross polarization and no higher order Floquet-Bloch (FB) harmonics; thereby, the reflected and transmitted fields can be expressed via
\begin{equation}
\label{Eq:EyScat}
    \begin{aligned}
        E_{y}^{\mathrm{ref}}\left(\vec{r}\right)&=r_{\mathrm{hom}}\!\left(\theta_0\right)E_0e^{-j\left(k_{x}x-k_{z,0}z\right)}&(z<0),\\
        E_{y}^{\mathrm{tran}}\left(\vec{r}\right)&=t_{\mathrm{hom}}\!\left(\theta_0\right)E_0e^{-j\left(k_{x}x+k_{z,0}z\right)}&(z>0),
    \end{aligned}
\end{equation}
where $r_{\mathrm{hom}}\!\left(\theta_0\right)$ and $t_{\mathrm{hom}}\!\left(\theta_0\right)$ are the reflection and transmission coefficients of the homogenized sheet in Fig.\ \ref{Fig:Theoretical_Config}(a), defined with respect to the $z=0$ reference plane.

We follow standard MS analysis \cite{Holloway2005,Holloway2009,Zaluski2016,Pfeiffer2016,Achouri2018,Momeni2021}, i.e., substitute the total fields, (\ref{Eq:EyInc}) and (\ref{Eq:EyScat}), in the GSTCs, (\ref{Eq:GSTCs})--(\ref{Eq:SusceptComponents}), and find the scattering coefficients in terms of the setup parameters (susceptibilities and angle of incidence). Similarly to \cite{Shaham2022,Shaham2023}, we find that these can be expressed as rational functions of the normal wavenumber,
\begin{equation}
\label{Eq:scatTE}
    \begin{aligned}
    r_{\mathrm{hom}}\!\left(\theta_0\right)&=\frac{r_{0}+r_{1}\widetilde{k}_{z,0}+r_{2}\widetilde{k}_{z,0}^{2}}{d_{0}+d_{1}\widetilde{k}_{z,0}+d_{2}\widetilde{k}_{z,0}^{2}+d_{3}\widetilde{k}_{z,0}^{3}},\\
    t_{\mathrm{hom}}\!\left(\theta_0\right)&=\frac{t_{1}\widetilde{k}_{z,0}+t_{3}\widetilde{k}_{z,0}^{3}}{d_{0}+d_{1}\widetilde{k}_{z,0}+d_{2}\widetilde{k}_{z,0}^{2}+d_{3}\widetilde{k}_{z,0}^{3}},
    \end{aligned}
\end{equation}
where
\begin{equation}
\label{Eq:Coeff}
    \begin{aligned}
        &\begin{aligned}
            &r_0=-2\left( \widetilde{\chi}_{\mathrm{ee}}^{yy}+\widetilde{\chi}_{\mathrm{mm}}^{zz}\right),
            &r_1=4\widetilde{\chi}_{\mathrm{em}}^{yx},\\
            &r_2=2\left( \widetilde{\chi}_{\mathrm{mm}}^{xx}+\widetilde{\chi}_{\mathrm{mm}}^{zz}\right),
        \end{aligned}\\
        &t_1=-j\left[\left(\widetilde{\chi}_{\mathrm{ee}}^{yy}+\widetilde{\chi}_{\mathrm{mm}}^{zz}\right) \widetilde{\chi}_{\mathrm{mm}}^{xx}+\left(\widetilde{\chi}_{\mathrm{em}}^{yx}\right)^2+4\right],\\
        &\begin{aligned}
            &t_3=j \widetilde{\chi}_{\mathrm{mm}}^{xx}\widetilde{\chi}_{\mathrm{mm}}^{zz},
            & d_0=2\left( \widetilde{\chi}_{\mathrm{ee}}^{yy}+\widetilde{\chi}_{\mathrm{mm}}^{zz}\right),
        \end{aligned}\\
        &d_{1}=j\left[ \left( \widetilde{\chi}_{\mathrm{ee}}^{yy}+\widetilde{\chi}_{\mathrm{mm}}^{zz}\right)\widetilde{\chi}_{\mathrm{mm}}^{xx}+\left(\widetilde{\chi}_{\mathrm{em}}^{yx}\right)^2-4\right],\\
        &\begin{aligned}
            &d_2=2\left( \widetilde{\chi}_{\mathrm{mm}}^{xx}-\widetilde{\chi}_{\mathrm{mm}}^{zz}\right),
            & d_3&=-j\widetilde{\chi}_{\mathrm{mm}}^{xx}\widetilde{\chi}_{\mathrm{mm}}^{zz}
        \end{aligned}
    \end{aligned}
\end{equation}
are the coefficients, and $\widetilde{\cdot}$ notation represents normalized dimensionless quantities with respect to the free-space wavenumber, i.e., $\widetilde{k}_{z,0}=k_{z,0}/k_0=\cos\theta_0$ and $\widetilde{\chi}=k_0\chi$.

The form of (\ref{Eq:scatTE}) and (\ref{Eq:Coeff}) is of great importance for several reasons. First, it is highly practical for engineering the angular behavior of the scattering coefficients, since it is analogous to standard frequency domain filters \cite{Matthei1980,Shaham2022,Shaham2023}: one may control the coefficients of the rational transfer functions and, hence, its functionality (all-pass, low-pass, etc.) and properties (zero and pole loci, cutoff angles, etc.), by tuning the susceptibilities accordingly. Second, it will soon play a key role in establishing the universal ties between the meta-atom and MS levels discussed in Sec.\ \ref{Sec:Introduction} (Sec.\ \ref{Subsec:Equivalence}).

We also notice the meta-atom-level nonlocal mechanism manifested in the angular dependence of the scattering coefficients of (\ref{Eq:scatTE}): as the susceptibilities in (\ref{Eq:Suscept}) are (approximately) local, it is the spatial derivatives that appear in Maxwell's equations and the GSTCs (\ref{Eq:GSTCs}) who are responsible for this spatial dispersion (also interpreted as weak nonlocality of the MS elements \cite{Asadchy2017PhD,Shastri2023,Shaham2023}). Physical meta-atom design schemes for the non-bianisotropic scenario ($\widetilde{\chi}_{\mathrm{em}}^{yx}=\widetilde{\chi}_{\mathrm{me}}^{xy}=0$) of (\ref{Eq:scatTE}) and (\ref{Eq:Coeff}) and dual derivations for the transverse magnetic (TM) polarization can be found in \cite{Shaham2022,Shaham2023}.

\subsection{Trilayered Admittance-Sheet Cascades (Metasurface Level)}
\label{Subsec:MSLevel}
Having established the meta-atom scattering formulation above, we shall now consider a well-known practical venue of HMS and OBMS realization at microwave frequencies: trilayered cascades of electrically polarizable admittance sheets compatible with PCB technology \cite{Pfeiffer2013Cascaded,Pfeiffer2013Millimeter,Epstein2016,Epstein2016OBMS,Epstein2016PRL} [Fig.\ \ref{Fig:Theoretical_Config}(b)]. The structure is composed of three layers, bottom (${z=-d}$), middle (${z=0}$), and top (${z=d}$), of surface admittance values $Y_{\mathrm{bot}}$, $Y_{\mathrm{mid}}$, and $Y_{\mathrm{top}}$, which we reserve as degrees of freedom yet to be tuned and realized in practice. The admittance $Y$ of a sheet accommodated on a certain plane, say ${z=z_0}$, is defined by relating the electric field therein ${E_y(z=z_0)}$ to the induced surface electric current, ${J_{\mathrm{s}y}(z=z_0)=H_{x}(z\to z_0^{+})-H_{x}(z\to z_0^{-})}$, via
\begin{equation}
\label{Eq:AdmitSheet}
    J_{\mathrm{s}y}(z=z_0)=YE_y(z=z_0).
\end{equation}
The admittance layers are separated by two dielectric substrates of relative permittivity $\epsilon_{\mathrm{r}}$, located at ${-d<z<0}$ and ${0<z<d}$, such that the overall thickness is $2d$.

Once again, we illuminate the composite from below by a $y$-polarized TE plane wave described via
\begin{equation}
\label{Eq:EyIncTri}
    E_y^{\mathrm{inc}}\left(\vec{r}\right)=E_0e^{-j\left[k_{x}x+k_{z,0}(z+d)\right]}\quad (z<-d),
\end{equation}
[Fig.\ \ref{Fig:Theoretical_Config}(b)], where $E_0$ is now defined as the amplitude at the reference plane of ${z=-d}$, and ${k_x=k_0\sin\theta_0}$ and $k_{z,0}=k_0\cos\theta_0$ are the wavevector components, as before. Similarly to Sec.\ \ref{Subsec:MALevel}, this transversely uniform TE MS configuration allows only fundamental FB harmonics in the spatial spectrum of the scattered fields, i.e., only plane waves of transverse wavenumber $k_{x}$ inherited from that of the incident wave in (\ref{Eq:EyIncTri}). Therefore, in the free-space regions of interest to us ($|z|>d$), the scattered [reflected $E_{y}^{\mathrm{ref}}\left(\vec{r}\right)$ and transmitted $E_{y}^{\mathrm{tran}}\left(\vec{r}\right)$] fields  can be expressed as
\begin{equation}
\label{Eq:EyCascade}
    \begin{aligned}
        E_{y}^{\mathrm{ref}}\left(\vec{r}\right)&=r_{\mathrm{tri}}\!\left(\theta_0\right)E_0e^{-j\left[k_{x}x-k_{z,0}\left(z+d\right)\right]}&(z<-d),\\
        E_{y}^{\mathrm{tran}}\left(\vec{r}\right)&=t_{\mathrm{tri}}\!\left(\theta_0\right)E_0e^{-j\left[k_{x}x+k_{z,0}\left(z-d\right)\right]}&(z>d),
    \end{aligned}
\end{equation}
where $r_{\mathrm{tri}}\!\left(\theta_0\right)$ and $t_{\mathrm{tri}}\!\left(\theta_0\right)$ are the reflection and transmission coefficients for the trilayered cascade [Fig.\ \ref{Fig:Theoretical_Config}(b)], defined with respect to the $z=\pm d$ reference planes.

Similarly to \cite{Monticone2013,Pfeiffer2013Cascaded,Pfeiffer2013Millimeter,PfeifferNano2014,Pfeiffer2014,Epstein2016,Epstein2016OBMS,Epstein2016PRL,Shaham2023}, we may represent the scattering problem via an equivalent transmission-line (TL) model \cite{Pozar2012}, as depicted in Fig.\ \ref{Fig:Theoretical_Config}(c): each of the free-space (dielectric) regions is described by an angularly dependent longitudinal propagation factor along the $z$ direction, ${\widetilde{k}_{z,0}=\cos\theta_0}$ [${\widetilde{k}_{z,\mathrm{d}}=(\epsilon_{\mathrm{r}}-\widetilde{k}_{x}^{2})^{1/2}=}{(\chi_{\mathrm{r}}+\widetilde{k}_{z,0}^{2})^{1/2}}$], and angularly dependent characteristic admittance, $\widetilde{Y}_0=\widetilde{k}_{z,0}$ ($\widetilde{Y}_{\mathrm{d}}=\widetilde{k}_{z,\mathrm{d}}$); the admittance sheets are equivalent to shunt loads of lumped admittance identical to their surface admittance ($\widetilde{Y}_{\mathrm{bot}}$, $\widetilde{Y}_{\mathrm{mid}}$, and $\widetilde{Y}_{\mathrm{top}}$). Hereafter, $\tilde{\cdot}$ denotes normalization of quantities into dimensionless values with respect to the wavenumber and admittance in free space, i.e., $\widetilde{k}_{z,0}=k_{z,0}/k_0$ ($\widetilde{k}_{z,\mathrm{d}}=k_{z,\mathrm{d}}/k_0$) and $\widetilde{Y}_0=\eta_0Y_0$ ($\widetilde{Y}_{\mathrm{d}}=\eta_0Y_{\mathrm{d}}$); $\chi_{\mathrm{r}}=\epsilon_{\mathrm{r}}-1$ is defined as the volumetric susceptibility of the dielectric slab.

Employing standard TL analysis \cite{Pozar2012} (Appendix \ref{App:TLderivation}), we express the scattering coefficients in terms of setup parameters,
\begin{equation}
\label{Eq:RTcascade}
    \!\!\!\begin{aligned}
        r_{\mathrm{tri}}\left(\theta_0\right)&\!=\!\frac{\widetilde{k}_{z,0}-\frac{h_{\mathrm{top}}^{+}l_{\mathrm{bot}}^{-}+h_{\mathrm{top}}^{-}\left(l_{\mathrm{mid}}^{+}+\widetilde{Y}_{\mathrm{bot}}l_{\mathrm{mid}}^{-}\right)}{jg^{-}h_{\mathrm{top}}^{+}+h_{\mathrm{top}}^{-}l_{\mathrm{mid}}^{-}}}{\widetilde{k}_{z,0}+\frac{h_{\mathrm{top}}^{+}l_{\mathrm{bot}}^{-}+h_{\mathrm{top}}^{-}\left(l_{\mathrm{mid}}^{+}+\widetilde{Y}_{\mathrm{bot}}l_{\mathrm{mid}}^{-}\right)}{jg^{-}h_{\mathrm{top}}^{+}+h_{\mathrm{top}}^{-}l_{\mathrm{mid}}^{-}}},\\
        t_{\mathrm{tri}}\left(\theta_0\right)&\!=\!\frac{\left[1+r\left(\theta_0\right)\right]\sec^{2}\left(k_{z,\mathrm{d}}d\right)}{h_{\mathrm{top}}^{-}l_{\mathrm{mid}}^{-}\!+\!jg^{-}h_{\mathrm{top}}^{+}},
    \end{aligned}
\end{equation}
where we define the angularly dependent auxiliary terms,
\begin{equation}
\label{Eq:AuxTerms}
    \begin{aligned}
        h_{\mathrm{top}}^{+}\!\left(\widetilde{k}_{z,0}\right)&=\widetilde{k}_{z,0}+\widetilde{Y}_{\mathrm{top}}+jg^{+}\!\!\left(\widetilde{k}_{z,0}\right),\\
        h_{\mathrm{top}}^{-}\!\left(\widetilde{k}_{z,0}\right)&=1+j\left(\widetilde{k}_{z,0}+\widetilde{Y}_{\mathrm{top}}\right)g^{-}\!\!\left(\widetilde{k}_{z,0}\right),\\
        l_{\mathrm{mid}}^{+}\!\left(\widetilde{k}_{z,0}\right)&=\widetilde{Y}_{\mathrm{mid}}+jg^{+}\!\!\left(\widetilde{k}_{z,0}\right),\\
        l_{\mathrm{mid}}^{-}\!\left(\widetilde{k}_{z,0}\right)&=1+j\widetilde{Y}_{\mathrm{mid}}g^{-}\!\!\left(\widetilde{k}_{z,0}\right),\\
        l_{\mathrm{bot}}^{-}\!\left(\widetilde{k}_{z,0}\right)&=1+j\widetilde{Y}_{\mathrm{bot}}g^{-}\!\!\left(\widetilde{k}_{z,0}\right),
    \end{aligned}
\end{equation}
and
\begin{equation}
\label{Eq:g}
    \begin{aligned}
        \!\!g^{\pm}\!\!\left(\widetilde{k}_{z,0}\right)&=\widetilde{Y}_{\mathrm{d}}^{\pm 1}\tan\left(k_{z,\mathrm{d}}d\right)\\
        &=\!\left(\!\chi_{\mathrm{r}}+\widetilde{k}_{z,0}^{2}\!\right)^{\pm 1/2}\!\tan\left[k_0 d\left(\chi_{\mathrm{r}}+\widetilde{k}_{z,0}^{2}\right)^{1/2}\right],
    \end{aligned}
\end{equation}
whose notation of angular dependence $\left(\widetilde{k}_{z,0}\right)$ is omitted in (\ref{Eq:RTcascade}) for brevity. At this stage, it is worth noting that similar TL scattering analyses of such multilayered MSs have been carried out in the past for the \emph{simple particular scenario of normal incidence} ($\widetilde{k}_{z,0}=1$) as a standard technique to engineer them \cite{Monticone2013,Pfeiffer2013Cascaded,Pfeiffer2013Millimeter,PfeifferNano2014,Pfeiffer2014,Epstein2016,Epstein2016OBMS,Epstein2016PRL}; herein, however, we purposely consider the \emph{entire angular dependence} of the scattering coefficients, to thus reveal new insights and practical implications, as follows.

Compared to the rational form obtained for the meta-atom setup in Sec.\ \ref{Subsec:MALevel}, we notice that the scattering coefficients of the cascade, (\ref{Eq:RTcascade})--(\ref{Eq:g}), depend transcendentally on the normal wavenumber, due to the tangent and secant functions that appear therein. This implies of an underlying nonlocal mechanism different from the previous meta-atom type: the trigonometric terms above are the consequence of infinitely many $e^{-jk_{z,\mathrm{d}}d}$ phase delays due to propagation across each of the dielectric substrates, which occurs between consecutive bounces off the partially reflective interfaces $z=0,\pm d$ \cite{Pozar2012,Balanis2012}. Since $\widetilde{k}_{z,\mathrm{d}}$ is determined from Snell's law (preservation of transverse momentum $k_{x}$), which is angularly dependent, it is clear that the nonlocal phenomenon at work is a combination of refraction into (and out of) the substrate and interference of multiple delayed reflections \cite{Gerken2003,Shastri2023}. This nonlocal phenomenon takes place in conjunction to another mechanism of the meta-atom type, due to the admittance relations defined in (\ref{Eq:AdmitSheet}); its spatial dispersion stems from the spatial derivatives in Maxwell's equations required to relate the tangential magnetic discontinuity [slightly before (\ref{Eq:AdmitSheet})] to the electric field therein \cite{Shaham2023}. At this stage, we are prepared to derive the following all-angle equivalence between the MS device in this section to its meta-atom counterpart in Sec.\ \ref{Subsec:MALevel}.

\subsection{All-Angle Equivalence Between Meta-Atom and MS Levels}
\label{Subsec:Equivalence}
Although the stack in Fig.\ \ref{Fig:Theoretical_Config}(b) bears a practical PCB means for MS realization \cite{Pfeiffer2013Cascaded,Pfeiffer2013Millimeter,Epstein2016,Epstein2016OBMS,Epstein2016PRL}, the angular dependence of the scattering coefficients off it, (\ref{Eq:RTcascade})--(\ref{Eq:g}), turns out to be cumbersome and nonintuitive. Namely, while the forward task of evaluating the scattering at all angles, given the setup parameters ($k_0d$, $\epsilon_{\mathrm{r}}$, $\widetilde{Y}_{\mathrm{bot}}$, $\widetilde{Y}_{\mathrm{mid}}$, and $\widetilde{Y}_{\mathrm{top}}$), is obvious, the inverse task of selecting such parameter values to achieve a given desired goal of an angular response may prove somewhat labyrinthine. A major reason to this difficulty is that the MS constitutive parameters ($k_0d$, $\epsilon_{\mathrm{r}}$, $\widetilde{Y}_{\mathrm{bot}}$, $\widetilde{Y}_{\mathrm{mid}}$, and $\widetilde{Y}_{\mathrm{top}}$) are intertwined with the angle-related parameter ($\widetilde{k}_{z,0}=\cos\theta_0$) in the complicated expressions of (\ref{Eq:RTcascade})--(\ref{Eq:g}).

In contrast, the GSTC-based rational scattering coefficients, (\ref{Eq:scatTE}) and (\ref{Eq:Coeff}), attained for the homogenized meta-atom scenario in Sec.\ {\ref{Subsec:MALevel}}, introduce a clear separation between such MS ($\widetilde{\chi}$) and angular ($\widetilde{k}_{z,0}$) parameters. Thus, they offer more intuition and greatly simplify the design procedure, as they yield closed-form expressions for the desired MS constituents (to be demonstrated in the next section). This appealing feature, along with the capability of the GSTCs (\ref{Eq:GSTCs}) to universally describe the operation of any well-behaved thin polarizable sheet \cite{Idemen1990,Tretyakov2003,Kuester2003,Achouri2018}, invokes the question ``can the MS stack in Fig.\ \ref{Fig:Theoretical_Config}(b) effectively realize the homogenized meta-atom sheet in Fig.\ \ref{Fig:Theoretical_Config}(a), (\ref{Eq:GSTCs})--(\ref{Eq:SusceptComponents}), for \emph{all} angles of incidence (provided that it is thin enough), as illustrated in Fig.\ \ref{Fig:Theoretical_Config}?''

A judicious course of action to answer this question would be first to apply Taylor's approximation to the transcendental functions in (\ref{Eq:RTcascade})--(\ref{Eq:g}), and thus approximate $r_{\mathrm{tri}}\left(\theta_0\right)$ and $t_{\mathrm{tri}}\left(\theta_0\right)$ into rational forms, as inspired by (\ref{Eq:scatTE}). The angle (or wavenumber) around which the Taylor approximation is derived marks the angle in which this approximation coincides precisely with the exact expression. Therefore, it should be generally chosen as the angle for (and around) which accuracy matters the most. As we have fundamentally revealed in \cite{Shaham2023} (for a more basic structure), we stress that the \emph{grazing} angle\footnote{We treat the scattering coefficients (\ref{Eq:RTcascade}) as continuous functions on the interval $0\leq k_{z,0}\leq 1$ ($-90^{\circ}\leq\theta_0\leq 90^{\circ}$); this holds for regular dielectric substrates, $\epsilon_{\mathrm{r}}>1$ ($\chi_{\mathrm{r}}>0$), with ${\widetilde{k}_{z,0}=0}$ as a removable discontinuity \cite{Shaham2023}.} $\theta_0=\pm 90^{\circ}$ ($\widetilde{k}_{z,0}=0$) is the proper choice for our purpose: at this angle, the wave immittance is singular (infinite impedance or zero admittance), such that even the slightest deviation in MS parameters may drastically affect the scattering coefficients near it; this is same reason why Fresnel reflection between two isotropic media inevitably approaches to total near grazing incidence. This inherent property and, thus, the requirement for uncompromised precision at $\widetilde{k}_{z,0}=0$, are well demonstrated in the context of the GHC \cite{Shaham2023}.

Hence, we evaluate the relevant Taylor approximations of the non-polynomial functions that appear in (\ref{Eq:RTcascade}) and (\ref{Eq:g}) around the grazing angle, $\theta_0=\pm 90^{\circ}$, i.e., $\widetilde{k}_{z,0}=0$:
\begin{equation}
\label{Eq:Taylor}
    \begin{aligned}
        &g^{+}\!\!\left(\widetilde{k}_{z,0}\right)\approx\; p+u\widetilde{k}_{z,0}^{2}+\mathcal{O}\left[\frac{1}{3}\left(k_0d\right)^3\widetilde{k}_{z,0}^{4}\right],\\
        &g^{-}\!\!\left(\widetilde{k}_{z,0}\right)\approx\; q+\mathcal{O}\left[\frac{1}{3}\left(k_0d\right)^3\right],\\
        &\begin{aligned}
        \sec^{2}\left(k_{z,\mathrm{d}}d\right)\approx&\,\left(1+pq\right)+q\left(2u-q\right)\widetilde{k}_{z,0}^{2}\\
        &\;+\mathcal{O}\left[-\frac{2}{3}\left(k_0d\right)^{4}\widetilde{k}_{z,0}^{4}\right],
        \end{aligned}
    \end{aligned}
\end{equation}
where
\begin{equation}
\label{Eq:pqus}
    \begin{aligned}
        &p=\chi_{\mathrm{r}}^{1/2}\tan\left(\chi_{\mathrm{r}}^{1/2}k_0d\right)=\mathcal{O}\left(\chi_{\mathrm{r}}k_0d\right),\\ &q=\chi_{\mathrm{r}}^{-1/2}\tan\left(\chi_{\mathrm{r}}^{1/2}k_0d\right)=\mathcal{O}\left(k_0d\right),\\
        &u=\frac{1}{2}\left[q+k_0d\sec^{2}\left(\chi_{\mathrm{r}}^{1/2}k_0d\right)\right]=\mathcal{O}\left(k_0d\right),\\
        &1+pq=\sec^{2}\left(\chi_{\mathrm{r}}^{1/2}k_0d\right)=\mathcal{O}\left(1\right),\\
        &2u-q=k_0d\sec^{2}\left(\chi_{\mathrm{r}}^{1/2}k_0d\right)=\mathcal{O}\left(k_0d\right)
    \end{aligned}
\end{equation}
are the coefficients, which depend solely on the substrate properties and frequency, namely, $k_0d$ and $\chi_{\mathrm{r}}$. The higher order terms in (\ref{Eq:Taylor}), 
$\mathcal{O}\left[\frac{1}{3}\left(k_0d\right)^3\widetilde{k}_{z,0}^{4}\right]$,  $\mathcal{O}\left[\frac{1}{3}\left(k_0d\right)^3\right]$, and $\mathcal{O}\left[-\frac{2}{3}\left(k_0d\right)^{4}\widetilde{k}_{z,0}^{4}\right]$, can be neglected [compared to the lower-order terms in (\ref{Eq:Taylor}) and (\ref{Eq:pqus})] for all angles ($-90^{\circ}<\theta_0<90^{\circ}$) when the substrates are electromagnetically thin with respect to a wavelength in free space, i.e.,  $\frac{2}{3}\left(k_0d\right)^{2}\ll$ 1.

Next, we substitute these approximations in (\ref{Eq:RTcascade}) and (\ref{Eq:AuxTerms}) and obtain the rational forms
\begin{equation}
\label{Eq:scatTEcascade}
    \begin{aligned}
    r_{\mathrm{tri}}\!\left(\theta_0\right)&=\frac{\rho_{0}+\rho_{1}\widetilde{k}_{z,0}+\rho_{2}\widetilde{k}_{z,0}^{2}}{\delta_{0}+\delta_{1}\widetilde{k}_{z,0}+\delta_{2}\widetilde{k}_{z,0}^{2}+\delta_{3}\widetilde{k}_{z,0}^{3}},\\
    t_{\mathrm{tri}}\!\left(\theta_0\right)&=\frac{\tau_{1}\widetilde{k}_{z,0}+\tau_{3}\widetilde{k}_{z,0}^{3}}{\delta_{0}+\delta_{1}\widetilde{k}_{z,0}+\delta_{2}\widetilde{k}_{z,0}^{2}+\delta_{3}\widetilde{k}_{z,0}^{3}},
    \end{aligned}
\end{equation}
where the coefficients
\begin{equation}
\label{Eq:TriCoeff}
    \begin{aligned}
        \rho_0&=\frac{j}{q}\left[\xi_{\mathrm{bot}}\left(\xi_{\mathrm{top}}\xi_{\mathrm{mid}}-2\xi\right)-\xi\left(\xi_{\mathrm{top}}-\xi_{\mathrm{bot}}\right)\right],\\
        \rho_{1}&=\xi_{\mathrm{mid}}\left(\xi_{\mathrm{top}}-\xi_{\mathrm{bot}}\right),\\
        \rho_{2}&=jq\xi_{\mathrm{mid}}-ju\left(\xi_{\mathrm{top}}+\xi_{\mathrm{bot}}\right),\\
        \tau_{1}&=2\xi,\quad
        \tau_{3}=2q\left(2u-q\right),\quad\delta_0=-\rho_0,\\ \delta_{1}&=\xi_{\mathrm{mid}}\left(\xi_{\mathrm{top}}+\xi_{\mathrm{bot}}\right)-2\xi,\\
        \delta_{2}&=jq\xi_{\mathrm{mid}}+ju\left(\xi_{\mathrm{top}}+\xi_{\mathrm{bot}}\right),\quad \delta_{3}=-2qu
    \end{aligned}
\end{equation}
depend on the MS constituent parameters and frequency, but not on angle, via the short-hand notations
\begin{equation}
\label{Eq:xi}
    \begin{aligned}
        \xi_{\mathrm{top}}&=1+jq\widetilde{Y}_{\mathrm{top}},\quad \xi_{\mathrm{bot}}=1+jq\widetilde{Y}_{\mathrm{bot}}\\
        \xi_{\mathrm{mid}}&=2+jq\widetilde{Y}_{\mathrm{mid}},\quad\xi=1+pq.
    \end{aligned}
\end{equation}
Remarkably, the rational form obtained for the scattering coefficients in (\ref{Eq:scatTEcascade}) is \emph{identical} to the one obtained at the meta-atom level in (\ref{Eq:scatTE}): specific particular forms of second- and third- order polynomials divided by a common third-order polynomial of the normal wavenumber, with coefficients that depend on the MS configuration.

This immediately leads us to the next step of deriving the equivalence between meta-atom and MS levels. To relate the realistic MS properties ($k_0d$, $\epsilon_{\mathrm{r}}$, $\widetilde{Y}_{\mathrm{bot}}$, $\widetilde{Y}_{\mathrm{mid}}$, and $\widetilde{Y}_{\mathrm{top}}$) to the equivalent meta-atom susceptibility values ($\widetilde{\chi}$), we enforce equality between the rational forms in (\ref{Eq:scatTE}) and (\ref{Eq:scatTEcascade}) \emph{for all angles} by setting their corresponding coefficients proportional to each other with a unique proportionality factor $\kappa$,
\begin{equation}
\label{Eq:Proportionality}
    \begin{aligned}
        &\left(r_0,r_1,r_2\right)=\kappa\left(\rho_0,\rho_1,\rho_2\right),\quad \left(t_1,t_3\right)=\kappa\left(\tau_1,\tau_3\right),\\
        &\left(d_0,d_1,d_2,d_3\right)=\kappa\left(\delta_0,\delta_1,\delta_2,\delta_3\right).
    \end{aligned}
\end{equation}
Indeed, we find that the only possible solution for this transformation reads $\kappa=-8j/\left[\xi_{\mathrm{mid}}\left(\xi_{\mathrm{top}}+\xi_{\mathrm{bot}}\right)\right]$ and
\begin{equation}
\label{Eq:Equivalence}
    \begin{aligned}
        &\widetilde{\chi}_{\mathrm{ee}}^{yy}=\frac{4\left[\xi_{\mathrm{bot}}\left(2\xi-\xi_{\mathrm{top}}\xi_{\mathrm{mid}}\right)+\xi\left(\xi_{\mathrm{top}}-\xi_{\mathrm{bot}}\right)\right]}{q\xi_{\mathrm{mid}}\left(\xi_{\mathrm{top}}+\xi_{\mathrm{bot}}\right)}+\frac{4u}{\xi_{\mathrm{mid}}}\\
        &\widetilde{\chi}_{\mathrm{mm}}^{xx}=\frac{4q}{\xi_{\mathrm{top}}+\xi_{\mathrm{bot}}},\quad\widetilde{\chi}_{\mathrm{mm}}^{zz}=-\frac{4u}{\xi_{\mathrm{mid}}},\\ &\widetilde{\chi}_{\mathrm{em}}^{yx}=-2j\frac{\xi_{\mathrm{top}}-\xi_{\mathrm{bot}}}{\xi_{\mathrm{top}}+\xi_{\mathrm{bot}}}.
    \end{aligned}
\end{equation}

Incredibly, substituting (\ref{Eq:Equivalence}) in (\ref{Eq:Coeff}) with the above solution for $\kappa$ and the expressions in (\ref{Eq:TriCoeff}) will satisfy \emph{all} the components of (\ref{Eq:Proportionality}), save from ${t_3/\kappa=2qu\neq\tau_3=2q(2u-q)}$. Nevertheless, this can be reconciled by expressing ${2u-q}\approx u+\mathcal{O}\left[\frac{1}{3}\chi_{\mathrm{r}}\left(k_0d\right)^{3}\right]$ and introducing another approximation in the spirit of the GSTCs \cite{Idemen1990,Tretyakov2003,Kuester2003,Achouri2018}: the MS must be thin not only with respect to a wavelength in free space [$\frac{2}{3}\left(k_0d\right)^{2}\ll 1$, as assumed slightly before (\ref{Eq:scatTEcascade})], but also compared to a dielectric wavelength as well, i.e., $\frac{1}{3}\chi_{\mathrm{r}}\left(k_0d\right)^{2}\ll 1$. This yields $t_3\approx\kappa\tau_3$, such that (\ref{Eq:Proportionality}) is entirely satisfied.

We have thus established the following universal link (Fig.\ \ref{Fig:Theoretical_Config}): \emph{for all the angles of incidence}, the space $-d<z<d$ occupied by the MS in Fig.\ \ref{Fig:Theoretical_Config}(b) can be effectively homogenized into the zero-thickness sheet of Fig.\ \ref{Fig:Theoretical_Config}(a) of surface susceptibilities prescribed in (\ref{Eq:Equivalence}), provided its thinness compared to free space and dielectric wavelengths,
\begin{equation}
\label{Eq:Approximations}
    \frac{2}{3}\left(k_0d\right)^{2}\ll 1\quad\mathrm{and}\quad  \frac{1}{3}\chi_{\mathrm{r}}\left(k_0d\right)^{2}\ll 1.
\end{equation}

Specifically, it rigorously emphasizes the nonintuitive yet inevitable appearance of \emph{normal} susceptibilities ($\chi_{\mathrm{mm}}^{zz}$) in the overall response. Early approaches, e.g., \cite{Monticone2013,Pfeiffer2013Cascaded,Pfeiffer2013Millimeter,PfeifferNano2014,Pfeiffer2014,Epstein2016,Epstein2016OBMS,Epstein2016PRL}, have associated the scattering off such composites solely with tangentially polarizable properties, i.e., $\chi_{\mathrm{ee}}^{yy}$, $\chi_{\mathrm{mm}}^{xx}$, and ${\chi_{\mathrm{em}}^{yx}=-\chi_{\mathrm{me}}^{xy}}$. While these responses, indeed, fully account for the MS functionality at normal incidence (due to the lack of normal magnetic field), they are not sufficient to describe it at other oblique angles; normal susceptibilities must be considered as well. In this regard, we have, in fact, retrieved the missing central part of the nonlocal puzzle, which, when adequately combined with its tangential counterparts, unlocks the possibility to tailor the MS response for \emph{all angles at once}. While, for a given set of substrate properties ($p$, $q$, and $u$), the MS composite introduces only three degrees of freedom ($\widetilde{Y}_{\mathrm{bot}}$, $\widetilde{Y}_{\mathrm{mid}}$, and $\widetilde{Y}_{\mathrm{top}}$) to control four meta-atom parameters ($\widetilde{\chi}_{\mathrm{ee}}^{yy}$, $\widetilde{\chi}_{\mathrm{mm}}^{xx}$, $\widetilde{\chi}_{\mathrm{mm}}^{zz}$, and $\widetilde{\chi}_{\mathrm{em}}^{yx}$), it still allows one to rigorously engineer a wide assortment of all-angle functionalities, as follows.

\section{Results and Discussion}

\subsection{Perfect Generalized Huygens' Metasurface Radomes}
\label{Subsec:Radome}

\subsubsection{Analytical Design} Our first demonstration of the profound formulation in Sec.\ \ref{Sec:Theory} concerns all-angle transmissive metasurfaces, which are beneficial for perfecting antenna radome applications \cite{Finley1956,Munk1971,Pelton1974,He2020,Goshen2024}. The following design extends our previous results in \cite{Shaham2023}, wherein a double-layered PCB cascade has been shown to strictly support only the grazing-angle Huygens' condition; herein, however, the triple-layered design would enable meeting the \emph{perfect} GHC, i.e., HC at both normal and grazing incidence scenarios \cite{ShahamEuCAP}. To this end, we impose $r_{\mathrm{tri}}\left(\theta_0\right)\equiv 0$ (assuming no loss); following (\ref{Eq:scatTEcascade}), this requires $\rho_0=\rho_1=\rho_2=0$, or, in view of the equivalence in Sec.\ \ref{Subsec:Equivalence}, $r_0=r_1=r_2=0$. This yields the GHC \cite{Shaham2023},
\begin{equation}
\label{Eq:GHC}
    -\widetilde{\chi}_{\mathrm{mm}}^{zz}=\widetilde{\chi}_{\mathrm{ee}}^{yy}=\widetilde{\chi}_{\mathrm{mm}}^{xx}=\widetilde{\chi}_{\mathrm{GHC}},\quad\widetilde{\chi}_{\mathrm{em}}^{yx}=0,
\end{equation}
where $\widetilde{\chi}_{\mathrm{GHC}}$ denotes the common susceptibility value, a degree of freedom to control the transmission phase \cite{Shaham2023}.

Subsequently, assuming a given set of substrate parameters and frequency, $k_0d$ and $\epsilon_{\mathrm{r}}$, which produce given values of $p$, $q$, and $u$ via (\ref{Eq:pqus}), we substitute the requirements (\ref{Eq:GHC}) in (\ref{Eq:Equivalence}). We first note that the lack of bianisotropy, $\widetilde{\chi}_{\mathrm{em}}^{yx}=0$, results in the symmetry of the MS around the $z=0$ plane, i.e., 
\begin{equation}
\label{Eq:Symmetric}
    \xi_{\mathrm{top}}=\xi_{\mathrm{bot}}\Leftrightarrow \widetilde{Y}_{\mathrm{top}}=\widetilde{Y}_{\mathrm{bot}};
\end{equation}
this is a well known property of (non-) bianisotropy in MSs \cite{Alaee2015,Epstein2016,Epstein2016OBMS,Achouri2020}. The next requirement, $-\widetilde{\chi}_{\mathrm{mm}}^{zz}=\widetilde{\chi}_{\mathrm{ee}}^{yy}$, which is, in fact, HC at the grazing angle \cite{Shaham2023} reads
\begin{equation}
\label{Eq:GrazingHC}
    \xi_{\mathrm{bot}}\left(\xi_{\mathrm{top}}\xi_{\mathrm{mid}}-2\xi\right)=\xi\left(\xi_{\mathrm{top}}-\xi_{\mathrm{bot}}\right).
\end{equation}
Substituting (\ref{Eq:Symmetric}) in (\ref{Eq:GrazingHC}), we obtain one possible solution,
\begin{equation}
\label{Eq:SymmAndGrazHC}
    \xi_{\mathrm{top}}\xi_{\mathrm{mid}}=2\xi.
\end{equation}

Next, we follow the second sub-condition in (\ref{Eq:GHC}), $\widetilde{\chi}_{\mathrm{mm}}^{xx}=\widetilde{\chi}_{\mathrm{ee}}^{yy}$, (HC at normal incidence \cite{Pfeiffer2013,Monticone2013,Pfeiffer2013Cascaded,Selvanayagam2013,Pfeiffer2013Millimeter,PfeifferNano2014,Wong2014,Epstein2016,Love1976,Jin2010,Decker2015,Shaham2023}); in our case, since we require the grazing-angle HC to hold as well, $-\widetilde{\chi}_{\mathrm{mm}}^{zz}=\widetilde{\chi}_{\mathrm{ee}}^{yy}$, (\ref{Eq:GrazingHC}), it would be mathematically equivalent (and easier) to demand $-\widetilde{\chi}_{\mathrm{mm}}^{zz}=\widetilde{\chi}_{\mathrm{mm}}^{xx}$ via (\ref{Eq:Equivalence}), subject to (\ref{Eq:Symmetric}),
\begin{equation}
\label{Eq:NormalHC}
    q\xi_{\mathrm{mid}}=2u\xi_{\mathrm{top}}.
\end{equation}
Finally, (\ref{Eq:SymmAndGrazHC}) and (\ref{Eq:NormalHC}) yield together a simple quadratic equation with two possible solutions,
\begin{equation}
    \begin{cases}
        \xi_{\mathrm{top}}^{\pm}=\xi_{\mathrm{bot}}^{\pm}=\pm\sqrt{q\xi/u},\\
        \xi_{\mathrm{mid}}^{\pm}=\pm 2\sqrt{u\xi/q},\\
        \widetilde{\chi}_{\mathrm{GHC}}^{\pm}=\pm 2\sqrt{qu/\xi}\approx\pm 2k_0d
    \end{cases}
    \label{Eq:GHCSolution}
\end{equation}
(distinguished from one another by the $+$ and $-$ superscripts). Following (\ref{Eq:xi}), we transform (\ref{Eq:GHCSolution}) into admittance form,
\begin{equation}
\label{Eq:GHC_Y}
    \begin{aligned}
        \widetilde{Y}_{\mathrm{top}}^{\pm}&=\widetilde{Y}_{\mathrm{bot}}^{\pm}=j\left(\frac{1}{q}\mp\sqrt{\frac{1+pq}{qu}}\right),\\
        \widetilde{Y}_{\mathrm{mid}}^{\pm}&=j\left[\frac{2}{q}\mp\frac{2}{q}\sqrt{\frac{u\left(1+pq\right)}{q}}\right].
    \end{aligned}
\end{equation}

We have thus obtained two closed-form solutions to perfectly realize generalized Huygens' MSs transparent at all angles with two different phase behaviors ($\widetilde{\chi}_{\mathrm{GHC}}^{\pm}\approx\pm 2k_0 d$). The ``$+$'' solution yields moderate admittance values, $\widetilde{Y}_{\mathrm{top}}^{+}=\widetilde{Y}_{\mathrm{bot}}^{+}\approx -j\frac{1}{3}\chi_{\mathrm{r}}k_0d$ and $\widetilde{Y}_{\mathrm{mid}}^{+}\approx-j\frac{4}{3}\chi_{\mathrm{r}}k_0d$, whereas the ``$-$'' one attains large values of $\widetilde{Y}_{\mathrm{top}}^{-}=\mathcal{O}\left(\frac{2j}{k_0d}\right)$ and $\widetilde{Y}_{\mathrm{mid}}^{-}=\mathcal{O}\left(\frac{4j}{k_0d}\right)$. Realization-wise, moderate admittance values typically allow one to set the working point away from resonance (to be discussed shortly); hence, we opt to validate the former, which we expect to feature wide bandwidth and low loss.

\begin{figure}[!t]
\centerline{\includegraphics[width=\columnwidth]{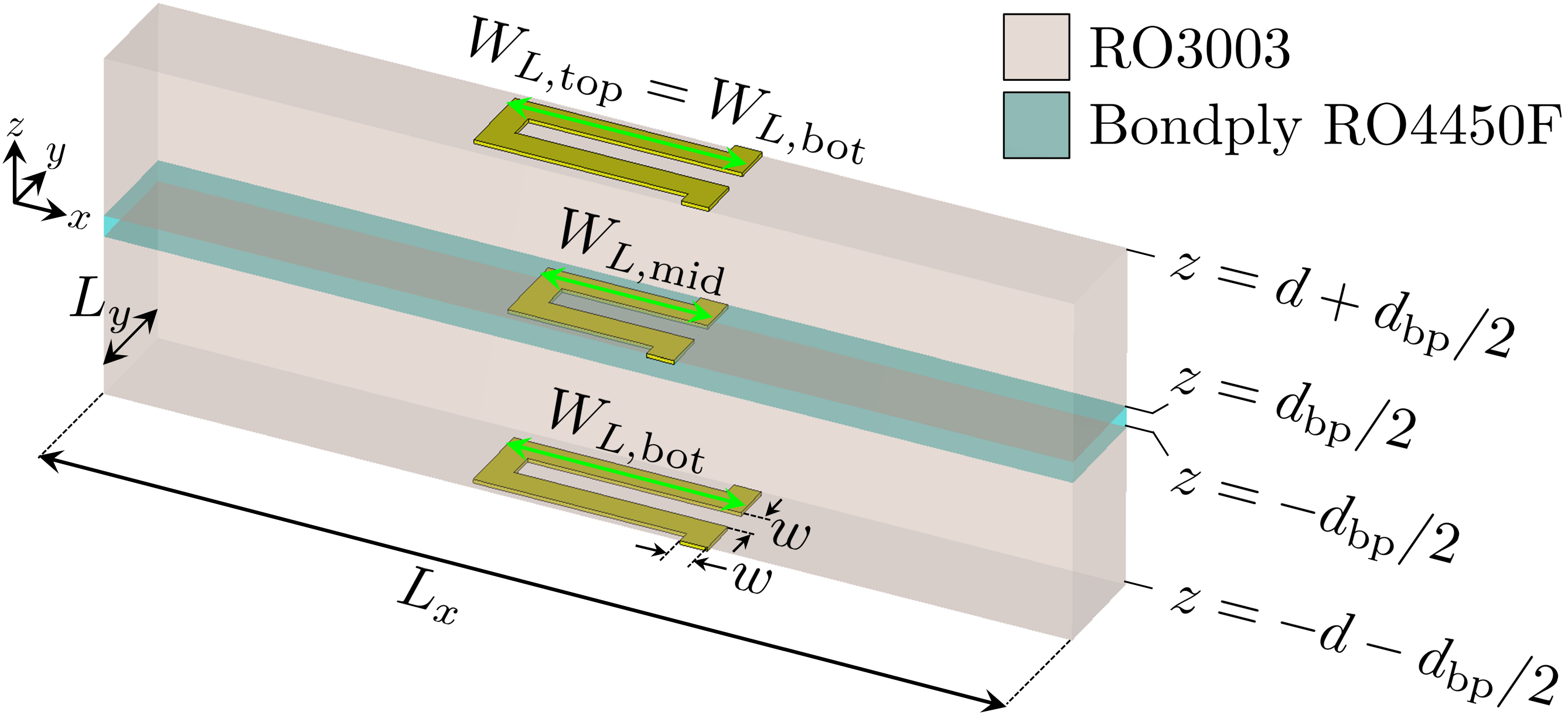}}
\caption{Practical PCB unit-cell configuration for the generalized Huygens' MS radome designed in Sec.\ \ref{Subsec:Radome} for $f=20$ GHz: two Rogers RO3003 dielectric substrates ($-d-d_{\mathrm{bp}}/2<z<-d_{\mathrm{bp}}/2$ and $d_{\mathrm{bp}}/2<z<d+d_{\mathrm{bp}}/2$) of permittivity $\epsilon_{\mathrm{r}}=3$, loss tangent $\tan\delta=0.001$, and thickness $d=30$ mil $=0.762$ mm $\approx0.0508\lambda_0$ each, bonded by a thin Rogers Bondply RO4450F of permittivity $\epsilon_{\mathrm{bp}}=3.52$, loss tangent $\tan\delta_{\mathrm{bp}}=0.004$, and thickness $d_{\mathrm{bp}}=4$ mil $=0.1016$ mm; printed 0.018-mm thick (0.5 oz/ft\textsuperscript{2}) copper meander-line realizations for the bottom ($z=-d-d_{\mathrm{bp}}/2$), top ($z=d+d_{\mathrm{bp}}/2$), and middle ($z=-d_{\mathrm{bp}}/2$) admittance sheets of common trace width $w=5$ mil $=0.127$ mm and respective meander widths, $W_{L,\mathrm{bot}}=W_{L,\mathrm{top}}$ and $W_{L,\mathrm{mid}}$, to be tuned in the following; the subwavelength unit-cell dimensions are $L_{x}=0.45$ mm $\approx0.3\lambda_0$ and $L_{y}=4w=20$ mil $=0.508$ mm.}
\label{Fig:GHMS_Config}
\end{figure}

\subsubsection{Full-Wave Realization and Validation}
\label{Subsubsec:GHC Full-wave}
To demonstrate this systematic design scheme, we set the frequency to ${f=20}$ GHz and select a commercial Rogers RO3003 substrate of $\epsilon_{\mathrm{r}}=3$ relative permittivity, $\tan\delta=0.001$ loss tangent, and standard  $d=30$ mil $=0.762$ mm $\approx 0.0508\lambda_0$ thickness (Fig.\ \ref{Fig:GHMS_Config}); $\lambda_0\approx 15$ mm is the wavelength at 20 GHz (free space). This set of parameters satisfies our thinness assumptions in (\ref{Eq:Approximations}): ${\frac{2}{3}\left(k_0 d\right)^{2}=\frac{1}{3}\chi_{\mathrm{r}}\left(k_0 d\right)^{2}\approx0.068\ll 1}$. We follow (\ref{Eq:pqus}) and find ${p\approx0.6861}$, ${q\approx0.3431}$, and ${u\approx0.3688}$, which lead, via (\ref{Eq:GHCSolution}) and (\ref{Eq:GHC_Y}), to the goal values of ${\widetilde{Y}_{\mathrm{top}}^{+}=\widetilde{Y}_{\mathrm{bot}}^{+}\approx-j0.2097}$, ${\widetilde{Y}_{\mathrm{mid}}^{+}\approx-j0.8888}$, and result in ${\widetilde{\chi}_{\mathrm{GHC}}^{+}\approx0.6401}$.

Next, to implement the inductive admittance values above via PCB technology, we propose a printed copper meander-line geometry for each of the layers (Fig.\ \ref{Fig:GHMS_Config}). The trace and gap widths are set to $w=5$ mil $=0.127$ mm, safely above the minimal feature size required by commercial PCB fabrication techniques; the copper layer thickness is 0.018 mm (standard $0.5$ oz/ft\textsuperscript{2} deposition). The strips are accommodated within subwavelength unit-cell dimensions of $L_{x}=4.5$ mm $\approx 0.3\lambda_0$ along $x$ and $L_{y}=4w\approx 0.034\lambda_0$ along $y$ (as dictated by the inclusion geometry). For manufacturing purposes, a thin commercial Rogers Bondply RO4450F of $\epsilon_\mathrm{bp}=3.52$ relative permittivity, $\tan\delta_{\mathrm{bp}}=0.004$ loss tangent, and $d_{\mathrm{bp}}=4$ mil $=0.1016$ mm thickness is inserted between the middle copper layer and the top dielectric substrate (Fig.\ \ref{Fig:GHMS_Config}).

Each of the meanders, in effect, serves as a distributed parallel LC load, whose resonant frequency can be respectively tuned through the meander widths, $W_{L,\mathrm{bot}}$, $W_{L,\mathrm{mid}}$, and $W_{L,\mathrm{top}}$ (Fig.\ \ref{Fig:GHMS_Config}); consequently, we may control the susceptance values at ${f=20}$ GHz through these widths \cite{Popov2019}, i.e., $\mathrm{Im}[\tilde{Y}_{\mathrm{bot}}\left(W_{L,\mathrm{bot}}\right)]$, $\mathrm{Im}[\tilde{Y}_{\mathrm{mid}}\left(W_{L,\mathrm{mid}}\right)]$, and $\mathrm{Im}[\tilde{Y}_{\mathrm{top}}\left(W_{L,\mathrm{top}}\right)]$. To find proper widths that accurately realize the above goal values ($\widetilde{Y}_{\mathrm{top}}^{+}=\widetilde{Y}_{\mathrm{bot}}^{+}\approx-j0.2097$, and $\widetilde{Y}_{\mathrm{mid}}^{+}\approx-j0.8888$), we proceed to construct a look-up table (LUT) that associates each relevant meander width $W_{L}$ to a corresponding susceptance value $\mathrm{Im}[\widetilde{Y}( W_{L})]$.

In general, various approaches to establish such LUTs (based mostly on full-wave simulations) have been reported in the past, e.g., \cite{Pfeiffer2013Cascaded,Pfeiffer2013Millimeter,PfeifferNano2014,Pfeiffer2014,Epstein2016,Epstein2016OBMS,Shaham2023}. Depending on the overall structure, the main considerations behind preferring one method over another are the accuracy of the tuning and the required number of parameter sweeps in full-wave simulations to achieve it (the less the better) \cite{Epstein2016}. In the following, we first briefly provide the characterization algorithm utilized to design the generalized Huygens' PCB MS in Fig.\ \ref{Fig:GHMS_Config} (Appendix \ref{App:Characterization}) and then explain the practical reasons for choosing it.

We first set $W_{L,\mathrm{top}}=W_{L,\mathrm{bot}}$ to ensure symmetry, i.e., $\widetilde{Y}_{\mathrm{top}}=\widetilde{Y}_{\mathrm{bot}}$, as instructed in (\ref{Eq:Symmetric}). We then construct a LUT for this common admittance value by modeling the structure in Fig.\ \ref{Fig:GHMS_Config}, without the middle copper trace (but including the bonding material), in the commercial full-wave solver ``CST Microwave Studio'' (CST): for a specific value of $W_{L,\mathrm{top}}$ under inspection, we retrieve reflection and transmission coefficients\footnote{The reference planes in the simulation are set to $z=\pm\left(d+d_{\mathrm{bp}}/2\right)$ and standard periodic boundary conditions are imposed.} at normal incidence, $r(0;W_{L,\mathrm{top}})$ and $t(0;W_{L,\mathrm{top}})$, from CST; we then substitute them in (\ref{Eq:CharYtop}) and (\ref{Eq:alpha}), Appendix \ref{App:Characterization}, to retrieve the susceptance value $\mathrm{Im}[\widetilde{Y}_\mathrm{top}( W_{L,\mathrm{top}})]$. By repeating this procedure for several values of $W_{L,\mathrm{top}}$ within a relevant range of interest, we find the LUT $\mathrm{Im}[\widetilde{Y}_{\mathrm{top}}(W_{L,\mathrm{top}})]$ and deduce that the value $W_{L,\mathrm{top}}=1.37$ mm achieves the required susceptance value above, $\mathrm{Im}(\widetilde{Y}_{\mathrm{top}})\approx -0.2097$, as instructed by the expression for $\widetilde{Y}_{\mathrm{top}}^{+}$ in (\ref{Eq:GHC_Y}).

Next, to find the suitable width of the middle meander $W_{L,\mathrm{mid}}$, we fix the top and bottom meander widths to the value obtained in the former stage ($W_{L,\mathrm{top}}=W_{L,\mathrm{bot}}=1.37$ mm) and reinstate the middle trace, of some given $W_{L,\mathrm{mid}}$ width under inspection, in the CST model as well. As before, we first simulate for the reflection and transmission coefficients off the complete structure at normal incidence, $r(0;W_{L,\mathrm{mid}})$ and $t(0;W_{L,\mathrm{mid}})$. We then substitute these new values in (\ref{Eq:alpha}) and (\ref{Eq:CharYmid}), Appendix \ref{App:Characterization}, and obtain the associated susceptance value $\mathrm{Im}[\widetilde{Y}_{\mathrm{mid}}\left(W_{L,\mathrm{mid}}\right)]$ for the width in question. Again, we apply this process for several values of $W_{L,\mathrm{mid}}$ and obtain the LUT $\mathrm{Im}[\widetilde{Y}_{\mathrm{mid}}\left(W_{L,\mathrm{mid}}\right)]$. Consecutively, we seek for the desired value $\mathrm{Im}(\widetilde{Y}_{\mathrm{mid}})\approx -0.8888$ in it, which we find to be realized via $W_{L,\mathrm{mid}}\approx0.634$ mm. We have, thereby, completed the main stage in designing such a generalized Huygens' MS radome of a trilayered PCB composite.

A note is in order regarding our choice of characterization technique. In our specific PCB design of a generalized Huygens' MS (Fig.\ \ref{Fig:GHMS_Config}), the electrical spacing between the layers, $k_0 d\sqrt{\epsilon_{\mathrm{r}}}\approx 0.55$, is taken deliberately small, as instructed by (\ref{Eq:Approximations}); this also leads to an appealing low-profile and lightweight device. Together with the relatively large (yet still sufficiently subwavelength) unit-cell period $L_{x}\approx 0.3\lambda_0$ required to support the above admittance values, this thin interlayer spacing may feature non-negligible nearfield effects. For this reason, we use the characterization methodology of (\ref{Eq:CharYtop})--(\ref{Eq:CharYmid}) above: on the one hand, it inherently provides partial yet reasonable consideration of nearfield effects and, accordingly, more accurate initial tuning; on the other hand, it still allows us to tune the dimensions $W_{L,\mathrm{top}}$ and $W_{L,\mathrm{mid}}$  \emph{separately}, similarly to \cite{PfeifferNano2014,Pfeiffer2014,Epstein2016OBMS}. This greatly reduces the number of required full-wave sweeps in the characterization process, similarly to \cite{Wong2014,Epstein2016}.

Having tuned the PCB MS according to the scheme above ($W_{L,\mathrm{top}}=W_{L,\mathrm{bot}}=1.37$ mm and $W_{L,\mathrm{mid}}=0.634$ mm), we now proceed to explore its properties. Specifically, in light of the equivalence in (\ref{Eq:Equivalence}), we expect this composite to virtually perform as an all-angle-transparent generalized Huygens' MS with susceptibility values of $-\widetilde{\chi}_{\mathrm{mm}}^{zz}=\widetilde{\chi}_{\mathrm{ee}}^{yy}=\widetilde{\chi}_{\mathrm{mm}}^{xx}=\widetilde{\chi}_{\mathrm{GHC}}^{+}\approx0.6401$, [as directly calculated in the beginning of this subsection from (\ref{Eq:GHCSolution})]. As the structure is symmetric around the $z=0$ plane (to a very good approximation, see Fig.\ \ref{Fig:GHMS_Config}), it is expected that the associated bianisotropic component $\widetilde{\chi}_{\mathrm{em}}^{yx}$ would practically vanish [as required in (\ref{Eq:GHC})]. That is, the only non-negligible susceptibility components that remain, according to our predictions above, are $\widetilde{\chi}_{\mathrm{ee}}^{yy}$, $\widetilde{\chi}_{\mathrm{mm}}^{xx}$, and $\widetilde{\chi}_{\mathrm{mm}}^{zz}$.

In this specific scenario, it is possible to systematically extract these corresponding susceptibility values exhibited by the configuration in question via the GSTC-based scheme proposed in \cite{Holloway2009,Zaluski2016}. It is important to emphasize that, like the GSTCs themselves, this scheme is \emph{universal} in the sense that it is valid for \emph{any} choice of microscopic realization, e.g., resonant loops and wires \cite{Shaham2022,Shaham2023}, provided that the MS manifests only the $\widetilde{\chi}_{\mathrm{ee}}^{yy}$, $\widetilde{\chi}_{\mathrm{mm}}^{xx}$, and $\widetilde{\chi}_{\mathrm{mm}}^{zz}$ susceptibilities mentioned above (and no other responses). Therefore, since this characterization methodology is not implementation-specific, it serves as an \emph{independent}, and, hence, \emph{objective} metric to probe our design and gain important insights about it \cite{Shaham2023}.

We employ this extraction method \cite{Holloway2009,Zaluski2016}, by simulating for the resultant reflection and transmission coefficients off the composite at normal incidence ($\theta_0=0$) and at one oblique angle, specifically $\theta_0=30^{\circ}$. Then, we substitute these results in the relevant equations in \cite{Shaham2023}, to obtain the susceptibility values. We find $\mathrm{Re}(\widetilde{\chi}_{\mathrm{ee}}^{yy})\approx0.6535$, $\mathrm{Re}(\widetilde{\chi}_{\mathrm{mm}}^{xx})\approx0.6627$, and $-\mathrm{Re}(\widetilde{\chi}_{\mathrm{mm}}^{zz})\approx 0.6133$, very close to our prediction for the common value $-\widetilde{\chi}_{\mathrm{mm}}^{zz}=\widetilde{\chi}_{\mathrm{ee}}^{yy}=\widetilde{\chi}_{\mathrm{mm}}^{xx}=\widetilde{\chi}_{\mathrm{GHC}}^{+}\approx0.6401$ (less than 4.2\% deviation). This strong agreement serves as a rigorous validation for the equivalence (\ref{Eq:Equivalence}) in Sec.\ \ref{Subsec:Equivalence}.

These results can be even further perfected by applying minute fine-tuning for the meander dimensions, to better account for nearfield effects (see previous discussion in this section). To achieve this, we first sweep the value of $W_{L,\mathrm{top}}$ in the presence of the middle wire of the current width $W_{L,\mathrm{mid}}$ and monitor $\widetilde{\chi}_{\mathrm{mm}}^{xx}$ while doing so. Next, according to the results of this sweep, we update the value of $W_{L,\mathrm{top}}$ to achieve the goal value $\mathrm{Re}(\widetilde{\chi}_{\mathrm{mm}}^{xx})=\widetilde{\chi}_{\mathrm{GHC}}\approx 0.6401$. Thereafter, we sweep $W_{L,\mathrm{mid}}$ and update its value to the one that achieves the grazing-angle Huygens' condition, $-\mathrm{Re}(\widetilde{\chi}_{\mathrm{mm}}^{zz})=\mathrm{Re}(\widetilde{\chi}_{\mathrm{ee}}^{yy})$. If necessary, this procedure can be repeated iteratively to improve accuracy; it converges very quickly (typically one or two sweeps) to the desired intersection of the three susceptibilities.

\begin{figure}[!t]
\centerline{\includegraphics[width=0.75\columnwidth]{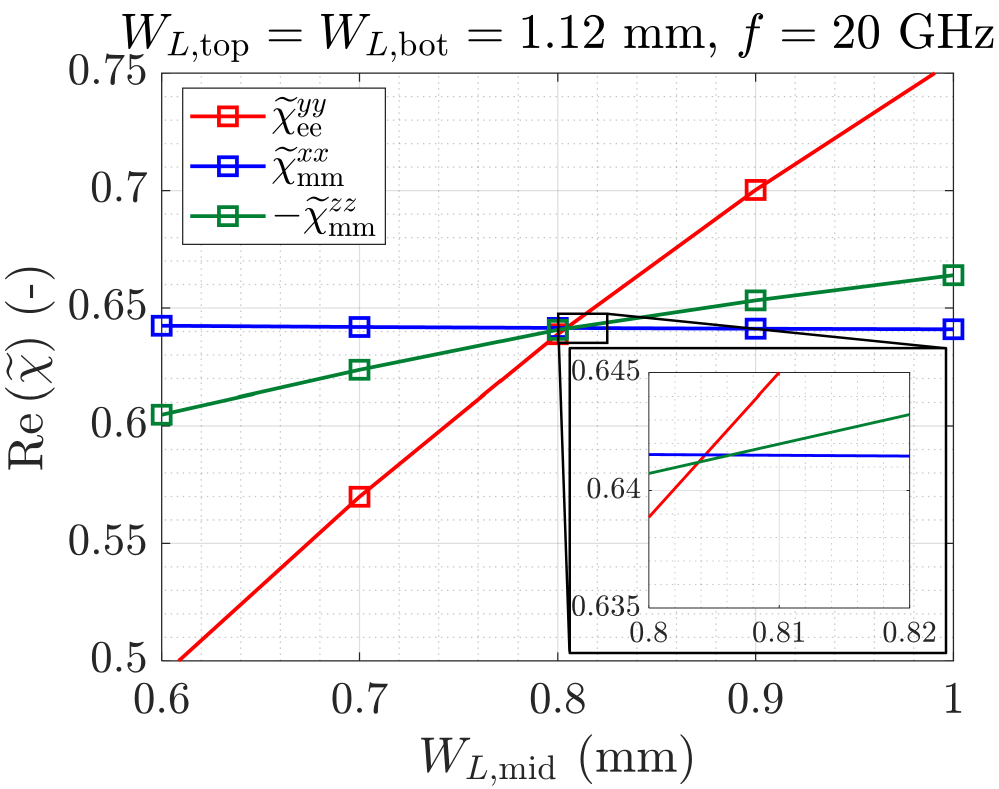}}
\caption{Meta-atom-level susceptibility characterization of the MS design in Fig.\ \ref{Fig:GHMS_Config} at $f=20$ GHz: real (reactive) part of the tangential electric ($\widetilde{\chi}_{\mathrm{ee}}^{yy}$, red solid lines with square markers), tangential magnetic ($\widetilde{\chi}_{\mathrm{mm}}^{xx}$, blue), and negative of the normal magnetic ($-\widetilde{\chi}_{\mathrm{mm}}^{zz}$, green) vs. middle meander width $W_{L,\mathrm{mid}}$ for $W_{L,\mathrm{top}}=W_{L,\mathrm{bot}}=1.12$ mm. Inset: close-up views near the approximate intersection of the three traces.}
\label{Fig:GHMS_Char}
\end{figure}

To demonstrate this, we show the final sweep of $W_{L,\mathrm{mid}}$ in Fig.\ \ref{Fig:GHMS_Char}, for which the value of $W_{L,\mathrm{top}}=W_{L,\mathrm{bot}}$ is set to 1.12 mm: establishing a LUT for the three susceptibility components at $f=20$ GHz, i.e., $\mathrm{Re}(\widetilde{\chi}_{\mathrm{ee}}^{yy})$ (red solid lines with square markers), $\mathrm{Re}(\widetilde{\chi}_{\mathrm{mm}}^{xx})$ (blue), and $-\mathrm{Re}(\widetilde{\chi}_{\mathrm{mm}}^{zz})$ (green), versus $W_{L,\mathrm{mid}}$, we find, remarkably, that they practically coincide at $-\widetilde{\chi}_{\mathrm{mm}}^{zz}\approx\widetilde{\chi}_{\mathrm{ee}}^{yy}\approx\widetilde{\chi}_{\mathrm{mm}}^{xx}\approx\widetilde{\chi}_{\mathrm{GHC}}\approx 0.6413$ for $W_{L,\mathrm{mid}}\approx0.804$ mm; this value is extremely close to the prediction in the beginning of this subsection, $\widetilde{\chi}_{\mathrm{GHC}}\approx0.6401$ [via (\ref{Eq:GHCSolution})]. Indeed, we deduce that the initial values obtained from the TL model, (\ref{Eq:CharYtop})--(\ref{Eq:CharYmid}), provide a good starting point (roughly 20\% deviation from the final widths, in this scenario) for fine-tuning the realistic PCB device under nearfield effects.

\begin{figure}[!t]
\centerline{\includegraphics[width=\columnwidth]{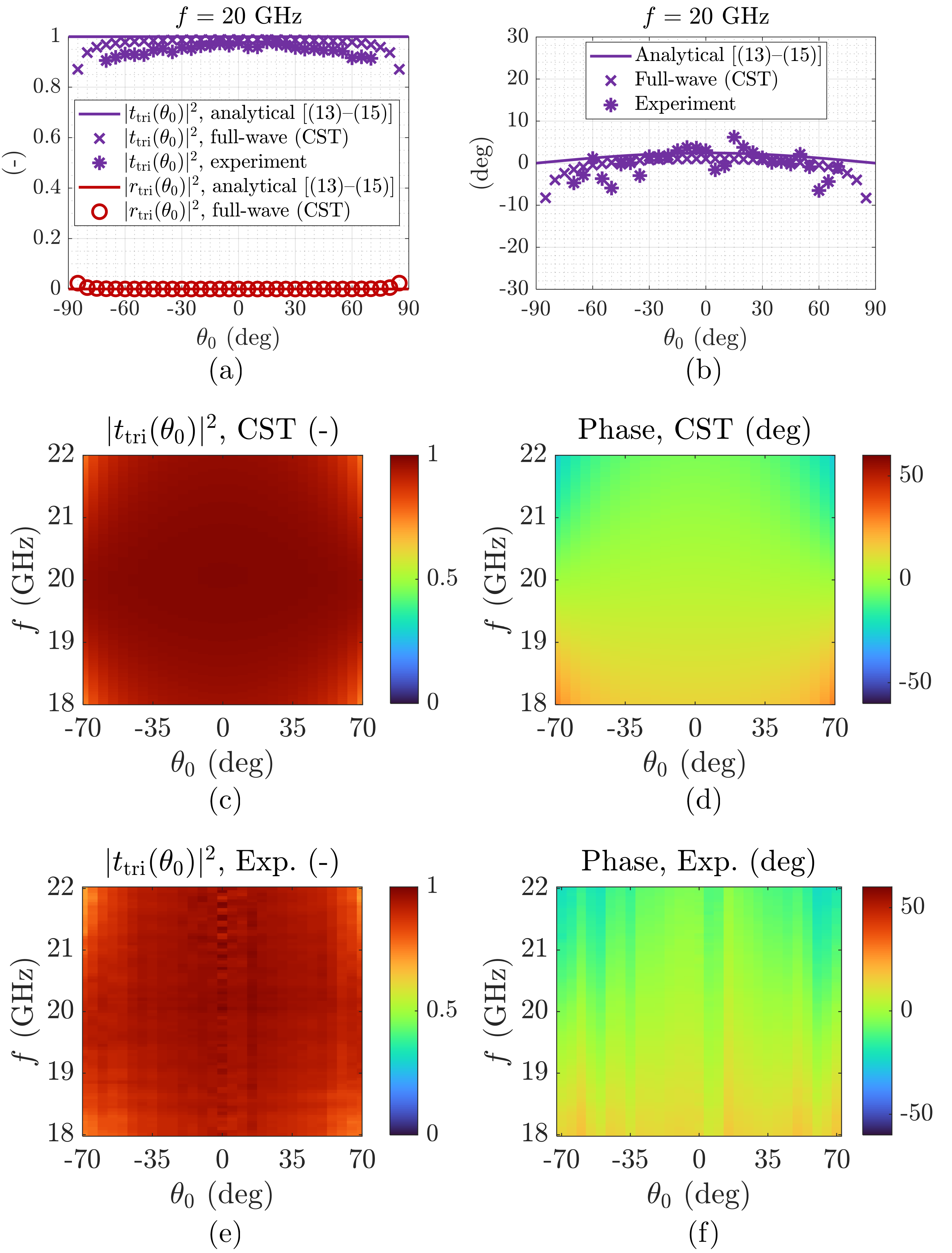}}
\caption{Analytical (solid lines) and full-wave (CST, markers) results of (a) transmittance $|t_{\mathrm{tri}}\left(\theta_0\right)|^{2}$ (purple, $\times$), reflectance $|r_{\mathrm{tri}}\left(\theta_0\right)|^{2}$ (red, $\circ$) and (b) transmission phase calibrated with respect to free-space propagation along the total MS width, $\angle t_{\mathrm{tri}}\left(\theta_0\right)+2k_0(d+d_{\mathrm{bp}}/2)$; corresponding experimental results (purple $*$ markers) of (a) transmittance and (b) calibrated transmission phase. Full-wave [CST, (c) and (d)] and corresponding experimental [(e) and (f)] frequency response of transmittance [(c) and (e)] and calibrated phase [(d) and (f)] plotted as a function of angle (ordinate) and frequency (abscissa).}
\label{Fig:GHMS_Results}
\end{figure}

We, therefore, finalize the design with $W_{L,\mathrm{top}}=W_{L,\mathrm{bot}}=1.12$ mm and $W_{L,\mathrm{mid}}= 0.804$ mm and inspect the theoretical [(\ref{Eq:RTcascade})--(\ref{Eq:g}), solid lines] and full-wave ($\times$ and $\circ$ markers) performance of its 20-GHz plane-wave reflectance (red, $\circ$ markers) and transmittance (purple, $\times$ markers) versus angle ($-85^{\circ}\leq\theta_0\leq 85^{\circ}$) in Fig.\ \ref{Fig:GHMS_Results}(a). Remarkably, the analytical reflectance practically vanishes, $|r_{\mathrm{tri}}\left(\theta_0\right)|^{2}<2.53\times 10^{-6}\approx -56$ dB, peaking at normal incidence and decreasing towards $0$ at $\theta_0\to\pm 90^{\circ}$; this desired behavior is expected as precise fulfillment of HC at grazing incidence (or, in general, the accuracy of any scattering condition at grazing incidence) is ingrained in our methodology (Sec.\ \ref{Subsec:Equivalence}). The transmittance complements the reflectance to unity (due to losslessness), i.e., $|t\left(\theta_0\right)|^{2}=1-|r\left(\theta_0\right)|^{2}\approx 1$.

In comparison, the full-wave reflectance attains its minimum at normal incidence, $|r\left(0\right)|^{2}\approx 19.2\times 10^{-6}\approx -47.17$ dB, and increases towards $|r\left(85^{\circ}\right)|^{2}\approx 2.3\%\approx -16.38$ dB, whereas the full-wave transmittance attains its near-unity maximum at normal incidence, $|t\left(0\right)|^{2}\approx 98.8\%$ and decreases towards $|t\left(85^{\circ}\right)|^{2}\approx 87.1\%$. Despite this minor performance degradation due to inevitable loss, the generalized Huygens' MS retains excellent transmittance above $93.7\%$ at the practical range of $-80^{\circ}\leq\theta_0\leq80^{\circ}$. Notably, very small values are attained by the reflection above, even for the challenging near-grazing angles of incidence---the transmittance is limited only by loss. This validates our theory above, showing that the GHC can be enforced in practical trilayered PCB configurations, also in the presence of realistic loss and nearfield effects.

Furthermore, to set the stage for comparison with experimental results, we calibrate the MS transmission phase $\angle t_{\mathrm{tri}}\left(\theta_0\right)$ with respect to the phase accumulated due to free-space propagation along the MS total thickness, $-k_0 \left(2d+d_{\mathrm{bp}}\right)\cos\theta_0$ (accounting for a reference measurement when the MS is removed), at 20 GHz. We present the analytical [(\ref{Eq:RTcascade})--(\ref{Eq:g}), solid lines] and full-wave ($\times$ markers) results of this calibrated phase, $\angle t_{\mathrm{tri}}\left(\theta_0\right)+k_0 \left(2d+d_{\mathrm{bp}}\right)\cos\theta_0$, versus angle of incidence in Fig.\ \ref{Fig:GHMS_Results}(b). The full-wave phase behavior also agrees very well with theory, slightly decreasing from $1.04^{\circ}$ at normal incidence to $-4^{\circ}$ at $\theta_0=\pm 80^{\circ}$, compared to $2.42^{\circ}$ and $0.42^{\circ}$ in theory. This extremely small deviation of MS transmission phase from its free-space propagation counterpart at all angles implies wavefront preservation capability \cite{Shaham2023}, which is many times attractive for radomes.

Additionally, we probe the full-wave frequency response of transmittance [Fig.\ \ref{Fig:GHMS_Results}(c)] and calibrated phase [Fig.\ \ref{Fig:GHMS_Results}(d)] in the band of $18$--$22$ GHz ($20\%$ fractional bandwidth around the center frequency), plotted as heatmaps of incidence angle as the ordinate and frequency as the abscissa. We find a relatively wide bandwidth operation, retaining $|t_{\mathrm{tri}}\left(\theta_0\right)|^{2}>86\%$ and $|\angle t_{\mathrm{tri}}\left(\theta_0\right)+k_0 \left(2d+d_{\mathrm{bp}}\right)\cos\theta_0|<21.84^{\circ}$ for $-60^{\circ}\leq\theta_0\leq 60^{\circ}$ in the above band, owing to the moderate admittance values discussed at the end of the former subsection. We thus have thoroughly validated our results in theory and simulation.

\begin{figure}[!t]
\centerline{\includegraphics[width=\columnwidth]{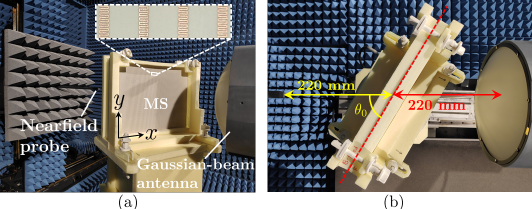}}
\caption{(a) Experimental setup consisting of a fabricated generalized Huygens' PCB MS radome, a transmitting Gaussian-beam antenna and a receiving nearfield probe; inset: detailed meander-line configuration of the top/bottom layer (periodic replication of the unit cell in Fig.\ \ref{Fig:GHMS_Config}).  (b) Top view of the measurement setup; the angle of incidence $\theta_{0}$ is set with the help of a goniometer with $0.5^{\circ}$ resolution.}
\label{Fig:GHMS_Exp_Setup}
\end{figure}

\subsubsection{Experimental Measurements}
To experimentally verify our results, a 9''$\times$12'' (${\approx15.24\lambda_0}$ along the meander $y$-direction and ${\approx20.32\lambda_0}$ along the $x$-direction) sample of the PCB-compatible design from Fig.\ \ref{Fig:GHMS_Config} was fabricated and characterized via measurements in the anechoic chamber at the Technion (Fig.\ \ref{Fig:GHMS_Exp_Setup}). The specimen was mounted on a rotatable foam holder, placed approximately at the focus of a Gaussian-beam antenna (Millitech Inc., GOA-42-S000094, focal distance of 196 mm $\approx 13\,\lambda_0$), illuminating the device under test (DUT) from a distance of $220$ mm with a quasi-planar wavefront.
A planar near-field measurement system (MVG/Orbit-FR) was utilized to record the forward scattering pattern, by scanning an area of $600$ mm $\times$ $500$ mm ($\approx 40\lambda_0\times33.3\lambda_0$) in the $x$ and $y$ directions, at a distance of $220$ mm from the center of the MS plane. The farfield pattern was then deduced via the equivalence principle \cite{Balanis2012}.

In each measurement, the foam holder was first rotated to set a specific angle of incidence in the range of $-70^{\circ}\leq\theta_{0}\leq70^{\circ}$ (limited by practical setup considerations). Next, an 18--22-GHz sweep was performed, followed by a reference measurement, for which the MS was removed and the Gaussian-beam antenna illuminated the same near-field scanning plane without it. The magnitude and phase of the forward farfield gain, which were calculated from the nearfield measurements and calibrated with respect to the reference measurements, represent the power transmittance $|t_{\mathrm{tri}}\left(\theta_0\right)|^{2}$ and the difference between the MS transmission phase $\angle t_{\mathrm{tri}}\left(\theta_0 \right)$ and the phase accumulated through free-space propagation along the total thickness of the absent MS $-k_{0}\left(2d+d_{\mathrm{bp}}\right)\cos\theta_0$, as before.

We present these results at $f=20$ GHz as purple $*$ markers in Figs.\ \ref{Fig:GHMS_Results}(a) and (b), in comparison to their formerly discussed analytical and full-wave counterparts. We observe very good agreement between experimental, full-wave, and analytical results, specifically, $|t_{\mathrm{tri}}\left(\theta_0\right)|^{2}>91.6\%$ and $|\angle t_{\mathrm{tri}}\left(\theta_0\right)+k_0\left(2d+d_{\mathrm{bp}}\right)\cos\theta_0|<6.5^{\circ}$ in $-60^{\circ}\leq\theta_0\leq 60^{\circ}$. We further juxtapose the experimental frequency response of transmittance [Fig.\ \ref{Fig:GHMS_Results}(e)] and calibrated transmission phase [Fig.\ \ref{Fig:GHMS_Results}(f)] to the previous full-wave results in the same format [Figs.\ \ref{Fig:GHMS_Results}(c) and (d)]. Indeed, good agreement is noted between experiment and theory, maintaining $|t_{\mathrm{tri}}\left(\theta_0\right)|^{2}>79.4\%$ and $|\angle t_{\mathrm{tri}}\left(\theta_0\right)+k_0\left(2d+d_{\mathrm{bp}}\right)\cos\theta_0|<22.76^{\circ}$ for the practical angular range $-60^{\circ}\leq\theta_0\leq 60^{\circ}$ in the above frequency band. These very good operation metrics and the overall results in this subsection validate our formalism and design scheme. Particularly, they show that the perfect GHC can be accurately realized in practice by a PCB cascade of three admittance-sheet layers, by the virtue of (\ref{Eq:GHC_Y}), thus enhancing the double-layered prototype in \cite{Shaham2023} to ultimate performance.

\subsection{All-Angle Perfect Magnetic Conductor}
\label{Subsec:PMC}
\subsubsection{Analytical Design}
To further stress the versatility of the framework developed in Sec.\ \ref{Sec:Theory}, we demonstrate yet another nonlocal design of an all-angle PMC plane via a PCB cascade of three layers. The PMC, dual of the perfect electric conductor (PEC), defines an abstract boundary condition of vanishing tangential magnetic field upon it. Unlike the PEC, which can be easily (yet, of course, non-ideally) realized by a good conductor at microwave frequencies, the PMC necessitates artificial MS implementation, e.g., \cite{Sievenpiper1999,Simovski2004,Feresidis2005,Monorchio2006,Hashemi2013}.
Since this boundary type is beneficial, e.g., as a ground plane for certain antennas \cite{Sievenpiper1999,Feresidis2005,Balanis2012}
or as a wall for a class of waveguides that support transverse-electric-magnetic (TEM) modes \cite{Kildal1990,Yang1999,Lier2010}, most of the above reports have resolved to stabilize its angular (and polarization) response. In this section, we harness our framework in Sec.\ \ref{Sec:Theory} to systematically design an all-angle PCB PMC MS (for the TE polarization).

Following its definition, a PMC plane exhibits unity reflection with zero phase and null transmittance for all angles. Specifically, for $\theta_0=\pm 90^{\circ}$ ($\widetilde{k}_{z,0}=0$), failure to comply with the grazing-angle HC, $-\widetilde{\chi}_{\mathrm{mm}}^{zz}=\widetilde{\chi}_{\mathrm{ee}}^{yy}$, inevitably sets the reflection coefficient therein to $r_0/d_0=-1$ [$\pi$ phase; see (\ref{Eq:scatTE}) and (\ref{Eq:Coeff})], manifesting undesired PEC behavior. The only possibility to avoid this situation is to impose the grazing-angle HC via (\ref{Eq:GrazingHC}), for which [see (\ref{Eq:scatTE}), (\ref{Eq:Coeff}), (\ref{Eq:scatTEcascade}), (\ref{Eq:Proportionality}), and (\ref{Eq:Equivalence})] 
\begin{equation}
\label{Eq:rPMC_Grazing}
    r_{\mathrm{tri}}\left(\theta_0\to 90^{\circ}\right)\approx\frac{\rho_1}{\delta_{1}}=\frac{r_1}{d_1}=\frac{4\widetilde{\chi}_{\mathrm{em}}^{yx}}{j\left[\left(\widetilde{\chi}_{\mathrm{em}}^{yx}\right)^{2}-4\right]}.
\end{equation}
Moreover, we ensure ${r_{\mathrm{tri}}\left(\theta_0\to 90^{\circ}\right)=1}$ by substituting this requirement in (\ref{Eq:rPMC_Grazing}) and obtaining a quadratic equation for the omega-bianisotropic component with a single solution,
\begin{equation}
\label{Eq:chiemyx}
    \widetilde{\chi}_{\mathrm{em}}^{yx}=-2j.
\end{equation}
This result was also found in \cite{Tiukuvaara2022}, albeit from a different set of considerations and without realization. In light of the fundamental property of bianisotropic MSs mentioned after (\ref{Eq:Symmetric}) \cite{Alaee2015,Epstein2016,Epstein2016OBMS,Achouri2020}, (\ref{Eq:chiemyx}) implies that the symmetry of the MS around $z=0$ must be broken, i.e., $\widetilde{Y}_{\mathrm{top}}\neq\widetilde{Y}_{\mathrm{bot}}$.

Following (\ref{Eq:Equivalence}), the only possibility to achieve (\ref{Eq:chiemyx}) is to set either $\xi_{\mathrm{bot}}=0$ or $\xi_{\mathrm{top}}\to\infty$. Substituting the former option in (\ref{Eq:GrazingHC}) results in $\xi_{\mathrm{top}}=\xi_{\mathrm{bot}}=0$, which, formally and physically speaking, leads to an ill-defined solution of a symmetric MS with ``$\widetilde{\chi}_{\mathrm{em}}^{yx}=0/0$'' bianisotropy. However, the latter solution,
\begin{equation}
\label{Eq:topPEC}
    \xi_{\mathrm{top}}\to\infty\quad\Leftrightarrow\quad\widetilde{Y}_{\mathrm{top}}\to\infty
\end{equation}
[see (\ref{Eq:xi})], achieves reasonable conditions when substituted in (\ref{Eq:GrazingHC}), i.e.,
\begin{equation}
\label{Eq:PMCGrazHC}
    \xi_{\mathrm{bot}}\xi_{\mathrm{mid}}=\xi.
\end{equation}
In practice, condition (\ref{Eq:topPEC}) can be satisfied by placing a conducting plane (ideally a PEC ground plane) as the top layer.

Therefore, we shall continue with (\ref{Eq:topPEC}) and (\ref{Eq:PMCGrazHC}), for which the effective susceptibilities in (\ref{Eq:Equivalence}) yield $\widetilde{\chi}_{\mathrm{mm}}^{xx}=0$, $\widetilde{\chi}_{\mathrm{ee}}^{yy}=-\widetilde{\chi}_{\mathrm{mm}}^{zz}=4u/\xi_{\mathrm{mid}}$, and (\ref{Eq:chiemyx}), such that the rational approximation of the reflection coefficient, (\ref{Eq:scatTE}) and (\ref{Eq:Coeff}), reduces into
\begin{equation}
\label{Eq:rPMC_Reduced}
    \bar{r}_{\mathrm{tri}}\left(\theta_0\right)\approx\frac{\xi_{\mathrm{mid}}-ju\widetilde{k}_{z,0}}{\xi_{\mathrm{mid}}+ju\widetilde{k}_{z,0}},
\end{equation}
where the top bar sign ( $\bar{\cdot}$ ) denotes the particular expression of $r_{\mathrm{tri}}\left(\theta_0\right)$ when (\ref{Eq:topPEC}) and (\ref{Eq:PMCGrazHC}) hold. Indeed, (\ref{Eq:rPMC_Reduced}) confirms that $\bar{r}_{\mathrm{tri}}\left(\theta_0\to 90^{\circ}\right)=1$. Furthermore, we recognize the rightmost expression in (\ref{Eq:rPMC_Reduced}) as the [1/1] Pad\'{e} approximant of $e^{-j2u\widetilde{k}_{z,0}/\xi_{\mathrm{mid}}}$ (p.\ 8 in \cite{Baker1996}),
\begin{equation}
\label{Eq:rPMCexp}
    \bar{r}_{\mathrm{tri}}\left(\theta_0\right)\approx e^{-j2u\widetilde{k}_{z,0}/\xi_{\mathrm{mid}}},
\end{equation}
which is applicable to sufficiently small arguments, e.g., $|u/\xi_{\mathrm{mid}}|\leq 0.4$ that ensures less than $\sim 2.2^{\circ}$ reflection phase deviation between (\ref{Eq:rPMC_Reduced}) and (\ref{Eq:rPMCexp}) for $-90^{\circ}<\theta_0<90^{\circ}$.

To understand this result, let us recall that $\bar{r}_{\mathrm{tri}}\left(\theta_0\right)$ [as a particular case of $r_{\mathrm{tri}}\left(\theta_0\right)$] is defined in (\ref{Eq:EyCascade}) with respect to the reference plane $z=-d$. We may shift the reference plane to another plane $z=z_0$ via (\ref{Eq:EyIncTri}) and (\ref{Eq:EyCascade}) \cite{Pozar2012} and find
\begin{equation}
\label{Eq:Shift}
    \begin{aligned}
        \bar{r}_{\mathrm{tri},z_0}(\theta_0)&=\frac{E_{y}^{\mathrm{ref}}\left(x,y,z_0\right)}{E_{y}^{\mathrm{inc}}\left(x,y,z_0\right)}\\
        &=\frac{E_{y}^{\mathrm{ref}}\left(x,y,-d\right)e^{+jk_0 \widetilde{k}_{z,0}\left(z_0+d\right)}}{E_{y}^{\mathrm{inc}}\left(x,y,-d\right)e^{-jk_0 \widetilde{k}_{z,0}\left(z_0+d\right)}}\\&=\bar{r}_{\mathrm{tri}}(\theta_0)e^{j2k_0 \widetilde{k}_{z,0}\left(z_0+d\right)}\\
        &\approx e^{j2\left[k_0\left(z_0+d\right)-u/\xi_{\mathrm{mid}}\right] \widetilde{k}_{z,0}},
    \end{aligned}
\end{equation}
where $\bar{r}_{\mathrm{tri},z_0}(\theta_0)$ denotes the reflection coefficient with respect to this new plane.
Specifically, if we select the plane
\begin{equation}
\label{Eq:zPMC}
    z_0=z_{\mathrm{PMC}}=-d+\frac{u}{k_0\xi_{\mathrm{mid}}},
\end{equation}
then (\ref{Eq:Shift}) states, \emph{for all angles}, that
\begin{equation}
\label{Eq:rPMCzPMC}
    \bar{r}_{\mathrm{tri},z_\mathrm{PMC}}\left(\theta_0\right)\approx 1.
\end{equation}

\begin{figure}[!t]
\centerline{\includegraphics[width=\columnwidth]{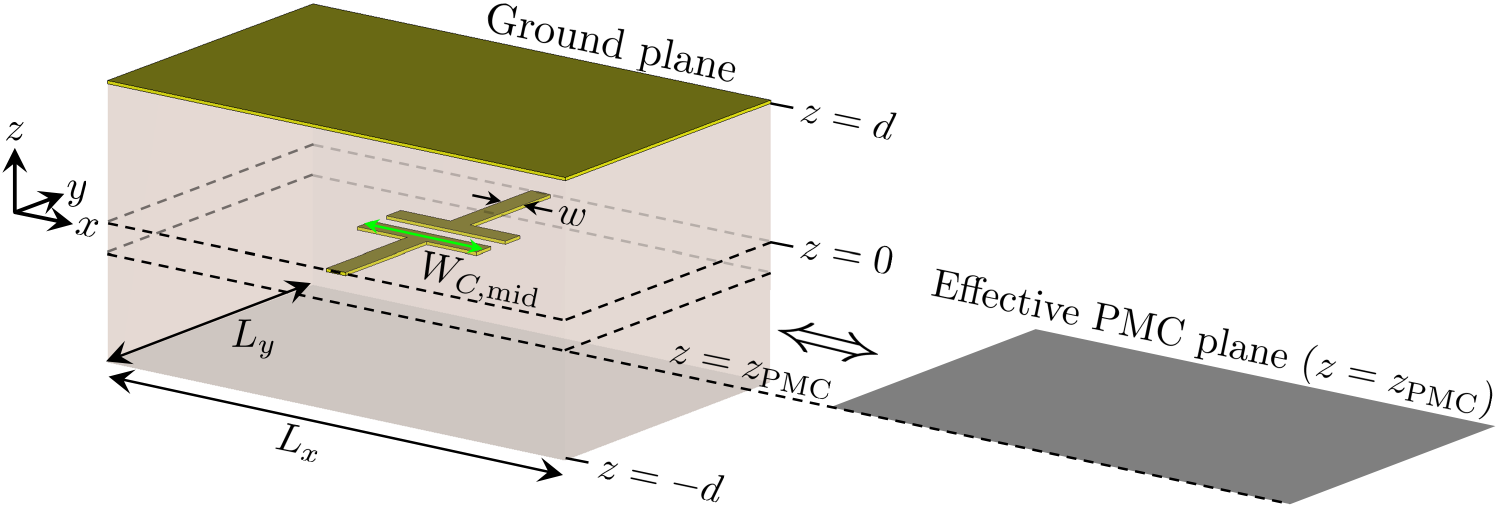}}
\caption{Left: a unit-cell of the PMC PCB design for $f=20$ GHz; the top layer ($z=d$) is grounded by a 0.018-mm thick copper plane ($\widetilde{Y}_{\mathrm{top}}\to\infty$); the middle layer ($z=0$) contains a capacitively loaded strip of $w=5$ mil $=0.127$ mm width and gap and $W_{C,\mathrm{mid}}$ capacitor width to be tuned; the bottom layer ($z=-d$) is left bare ($\widetilde{Y}_{\mathrm{bot}}=0$); the unit-cell size is $L_{x}=3$ mm $\approx0.2\lambda_0$ along $x$ and $L_{y}=1.8$ mm $\approx0.12\lambda_0$ along $y$. Right: an effective PMC plane emulated by the left structure at $z=z_{\mathrm{PMC}}$ defined in (\ref{Eq:zPMC}), slightly below the $z=0$ plane for the chosen parameters in Sec.\ \ref{Subsec:PMC}.}
\label{Fig:PMC_Config}
\end{figure}

We have thus effectively devised an all-angle PMC plane at $z=z_{\mathrm{PMC}}$ (Fig.\ \ref{Fig:PMC_Config}). That is, provided that the top layer is grounded (\ref{Eq:topPEC}) and that the middle and bottom layers satisfy (\ref{Eq:PMCGrazHC}), a PMC plane is effectively emulated at $z=z_{\mathrm{PMC}}$, whose value is given in (\ref{Eq:zPMC}). The value of $\xi_{\mathrm{mid}}$ (and, correspondingly, $\xi_{\mathrm{bot}}$) can be tuned to control the location of the PMC plane. The larger this value, the better the accuracy of the approximations in (\ref{Eq:rPMCexp}), (\ref{Eq:Shift}), and (\ref{Eq:rPMCzPMC}) for small angles of incidence $\theta_0$ as well (following the properties of the Pad\'{e} approximant), and the smaller the shift of the PMC plane (\ref{Eq:zPMC}) from the $z=-d$ reference plane; on the other hand, performance metrics, such as bandwidth and loss may be generally affected by this choice, especially for typical resonant realizations.

\subsubsection{Full-Wave Realization and Validation}
For the demonstration herein, we choose to study a simple yet practical case where the bottom layer is removed, i.e., $\widetilde{Y}_{\mathrm{bot}}=0$. This would greatly facilitate the design procedure, as it will spare us the trouble of co-tuning two nearfield coupled layers in proximity to resonance. By virtue of (\ref{Eq:xi}), (\ref{Eq:topPEC}), and (\ref{Eq:PMCGrazHC}), we obtain
\begin{equation}
\label{Eq:PMC_NoBot}
    \begin{gathered}
        \xi_{\mathrm{bot}}=1,\quad \xi_{\mathrm{mid}}=\xi,\quad\tilde{Y}_{\mathrm{top}}\to\infty, \quad\widetilde{Y}_{\mathrm{bot}}=0,\\
        \widetilde{Y}_{\mathrm{mid}}=\frac{j}{q}\left(2-\xi\right)\approx\frac{j}{q},\quad z_{\mathrm{PMC}}=-d+\frac{u}{k_0\xi}\approx 0.
    \end{gathered}
\end{equation}

Again, we set the frequency $f=20$ GHz ($\lambda_0=15$ mm) and consider a Rogers RO3003 substrate ($\epsilon_{\mathrm{r}}=3$ and $\tan\delta=0.001$) of total thickness $2d=60$ mil $=1.524$ mm $\approx 0.102\lambda_0$ (Fig.\ \ref{Fig:PMC_Config}), for which, following (\ref{Eq:pqus}) and (\ref{Eq:xi}), $p\approx0.6861$, $q\approx0.3431$, $u\approx0.3688$, and $\xi\approx 1.2354$. Next, we follow (\ref{Eq:PMC_NoBot}) and obtain the goal admittance value, $\widetilde{Y}_{\mathrm{mid}}\approx j2.2288$, and the expected effective PMC plane $z_{\mathrm{PMC}}\approx-0.0498$ mm $\approx-1.9606$ mil $\approx -0.0033\lambda_0$ (Fig.\ \ref{Fig:PMC_Config}).

To realize this capacitive admittance value via PCB technology, we propose a $y$-directed capacitively loaded printed strip of practical $w=5$ mil $=0.127$ mm width and gap, contained in subwavelength unit-cell dimensions of $L_{x}=3$ mm $\approx 0.2\lambda_0$ along $x$ and $L_{y}=1.8$ mm $\approx 0.12\lambda_0$ along $y$ (Fig.\ \ref{Fig:PMC_Config}). By tuning the capacitor width $W_{C,\mathrm{mid}}$, we may control the serial LC resonant frequency attributed to this inclusion and, hence, its susceptance value $\mathrm{Im}\left[\widetilde{Y}_{\mathrm{mid}}\left(W_{C,\mathrm{mid}}\right)\right]$ at $f=20$ GHz. The top layer is grounded by a copper plane to achieve $\widetilde{Y}_{\mathrm{top}}\to \infty$, and the bottom facet is left bare to satisfy $\widetilde{Y}_{\mathrm{bot}}=0$; no bonding layer is considered for simplicity.

\begin{figure}[!t]
\centerline{\includegraphics[width=0.8\columnwidth]{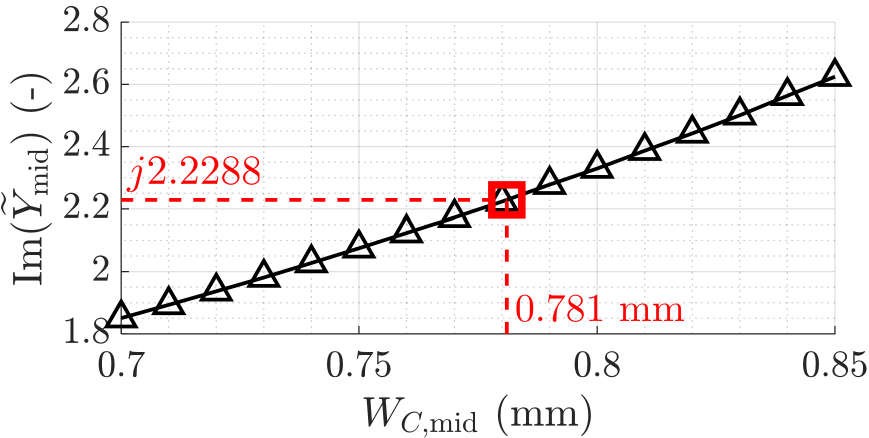}}
\caption{Look-up table (LUT) for the unit-cell configuration in Fig.\ \ref{Fig:PMC_Config}: surface susceptance of the middle layer $\mathrm{Im}[\widetilde{Y}_{\mathrm{mid}}\left(W_{C,\mathrm{mid}}\right)]$ versus capacitor width $W_{C,\mathrm{mid}}$, as characterized via (\ref{Eq:YmidPMC}) in CST; red dashed lines and square marker mark the desired susceptance value $\widetilde{Y}_{\mathrm{mid}}=j2.2288$ that occurs for $W_{C,\mathrm{mid}}=0.781$ mm.}
\label{Fig:PMC_LUT}
\end{figure}

Next, we establish a LUT that assigns 
a corresponding admittance $\widetilde{Y}_{\mathrm{mid}}\left(W_{C,\mathrm{mid}}\right)$ value to a given width $W_{C,\mathrm{mid}}$. To this end, we model the unit-cell of Fig.\ \ref{Fig:PMC_Config} in CST, with a specific $W_{C,\mathrm{mid}}$ value under inspection, impose periodic boundary conditions, and simulate for the plane-wave reflection coefficient off it, $\bar{r}_{\mathrm{tri}}\left(\theta_0;W_{C,\mathrm{mid}}\right)$, at a certain known interrogation angle $\theta_{\mathrm{0}}$. To construct a characterization formula that would enable us to extract the admittance value $\widetilde{Y}_{\mathrm{mid}}$ from the simulated reflection coefficient $\bar{r}_{\mathrm{tri}}\left(\theta_0;W_{C,\mathrm{mid}}\right)$, we substitute $\widetilde{Y}_{\mathrm{top}}\to\infty$ and $\widetilde{Y}_{\mathrm{bot}}=0$ in the exact expression for $r_{\mathrm{tri}}\left(\theta_0\right)$ [(\ref{Eq:RTcascade})--(\ref{Eq:g})], and obtain a closed-form expression for $\bar{r}_{\mathrm{tri}}\left(\theta_0\right)$ in terms of $\widetilde{Y}_{\mathrm{mid}}$ and the other setup parameters; then, we eliminate $\widetilde{Y}_{\mathrm{mid}}$ and find
\begin{equation}
\label{Eq:YmidPMC}
    \begin{aligned}
        &\mathrm{Im}\left[\widetilde{Y}_{\mathrm{mid}}\right]\\
        &=\mathrm{Im}\left[\frac{j}{g^{-}\!\!\left(\widetilde{k}_{z,0}\right)}+\frac{jg^{+}\!\!\left(\widetilde{k}_{z,0}\right)-\frac{1-\bar{r}_{\mathrm{tri}}\left(\theta_0\right)}{1+\bar{r}_{\mathrm{tri}}\left(\theta_0\right)}\widetilde{k}_{z,0}}{jg^{-}\!\!\left(\widetilde{k}_{z,0}\right)\!\!\frac{1-\bar{r}_{\mathrm{tri}}\left(\theta_0\right)}{1+\bar{r}_{\mathrm{tri}}\left(\theta_0\right)}\widetilde{k}_{z,0}-1}\right].
    \end{aligned}
\end{equation}

Since the angle of interrogation $\theta_0$ and the substrate properties are known in advance, we may find ${\widetilde{k}_{z,0}=\cos\theta_0}$ and $g^{\pm}\!\!\left(\widetilde{k}_{z,0}\right)$ via ($\ref{Eq:g}$) and substitute them, along with the simulated reflection coefficient $\bar{r}_{\mathrm{tri}}\left(\theta_0;W_{C,\mathrm{mid}}\right)$, in (\ref{Eq:YmidPMC}) to obtain $\widetilde{Y}_{\mathrm{mid}}\left(W_{C,\mathrm{mid}}\right)$. In principle, according to the homogenized admittance sheet model in Sec.\ \ref{Subsec:MSLevel}, these characterization results should not depend on the choice of interrogation angle $\theta_0$; however, in practice, the loaded-line realization in Fig.\ \ref{Fig:PMC_Config} would exhibit small angular dependence due to minor nearfield effects (see Sec. \ref{Subsubsec:GHC Full-wave}). Thus, as implied by the key foundations of our model, detailed slightly before (\ref{Eq:Taylor}), it would be preferable to set the near-grazing angle $\theta_0=85^{\circ}$ for this task. We
repeat this process for a relevant range of $W_{C,\mathrm{mid}}$ and thus arrive at the LUT, $\mathrm{Im}[\widetilde{Y}_{\mathrm{mid}}\left(W_{C,\mathrm{mid}}\right)]$, shown as triangle markers in Fig.\ \ref{Fig:PMC_LUT}. Subsequently, to set the aforementioned value of $\mathrm{Im}\left(\widetilde{Y}_{\mathrm{mid}}\right)\approx 2.2288$, we fetch the corresponding width $W_{C,\mathrm{mid}}=0.781$ mm from the LUT (red dashed lines and a square marker in Fig.\ \ref{Fig:PMC_LUT}) and fix it.

\begin{figure}[!t]
\centerline{\includegraphics[width=0.95\columnwidth]{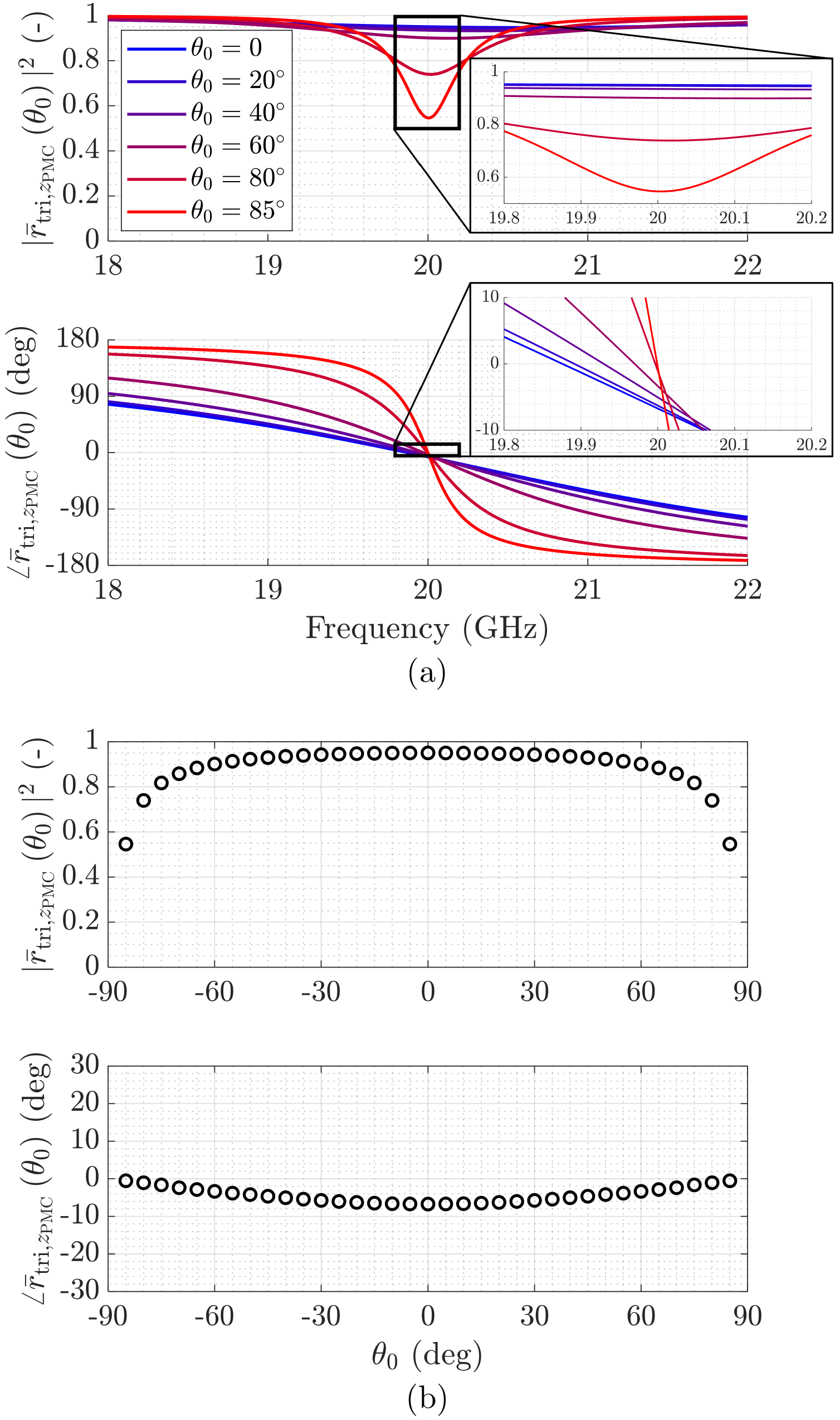}}
\caption{(a) Full-wave frequency response of reflectance ($|\bar{r}_{\mathrm{tri},z_{\mathrm{PMC}}}\left(\theta_0\right)|^{2}$, top subplot) and reflection phase at $z=z_{\mathrm{PMC}}$ ($\angle \bar{r}_{\mathrm{tri},z_{\mathrm{PMC}}}\left(\theta_0\right)$, bottom subplot) off the finalized PMC structure in Fig.\ \ref{Fig:PMC_Config}, with $W_{\mathrm{mid}}=0.781$ mm, for several selected angles, from $\theta_0=0$ (blue) to $\theta_0=85^{\circ}$ (red), passing through $\theta_0=20^{\circ}$, $40^{\circ}$, $60^{\circ}$, and $80^{\circ}$ (gradually changing colors from blue to red, see legend); insets: close-up around the center frequency $f=20$ GHz. (b) A cut of (a) at $f=20$ GHz showing the reflectance ($\circ$ markers in the top subfigure) and reflection phase ($\circ$ markers, bottom subfigure) vs.\ angle.}
\label{Fig:PMC_Results}
\end{figure}

We examine the performance of this finalized PMC MS in CST, by enforcing periodic boundary conditions and simulating for the TE plane-wave reflection $\bar{r}_{\mathrm{tri},z_{\mathrm{PMC}}}\left(\theta_0\right)$. The top subplot of Fig.\ \ref{Fig:PMC_Results}(a) shows the reflectance $|r_{\mathrm{tri},z_{\mathrm{PMC}}}\left(\theta_0\right)|^{2}$ versus frequency in the 18--22 GHz band for several angles from $\theta_0=0$ (blue trace) to $\theta_0=85^{\circ}$ (red), passing through $\theta_0=20^{\circ}$, $40^{\circ}$, $60^{\circ}$, and $80^{\circ}$ (gradually changing colors from blue to red). We observe an absorptive resonance (reflectance dip) gradually forming at 20 GHz (the goal frequency) as the angle is increased towards grazing incidence, $\theta_0\to90^{\circ}$, starting from a relatively flat near-unity frequency response at normal incidence, $\theta_0=0$.

The corresponding reflection phase $\angle \bar{r}_{\mathrm{tri},z_{\mathrm{PMC}}}\left(\theta_0\right)$ (bottom subplot in Fig.\ \ref{Fig:PMC_Results}) shows gradual transition from $180^{\circ}$ to $-180^{\circ}$ as the frequency is increased. Moreover, as the angle of incidence increases, this transition occurs more rapidly, approaching a step-like changeover at the limit of grazing incidence $\theta_0\to 90^{\circ}$, such that the desired zero phase (PMC) condition, $\angle \bar{r}_{\mathrm{tri},z_{\mathrm{PMC}}}\left(\theta_0\right)$, is approximately achieved at $f=20$ GHz, especially for near-grazing angles. In other words, to a good approximation, the traces in the bottom subplot crowd together near zero reflection phase at $f=20$ GHz and  (see inset), very close to the desired operation.

This can also be observed in the constant-frequency cut at $f=20$ GHz presented in Fig.\ \ref{Fig:PMC_Results}(b): the reflectance (top subplot, black $\circ$ markers) peaks at normal incidence, $|r_{\mathrm{tri},z_{\mathrm{PMC}}}\left(0\right)|^{2}\approx 95\%$, and plummets toward zero at the absorptive resonant grazing angle. Nevertheless, good performance of $|r_{\mathrm{tri},z_{\mathrm{PMC}}}\left(\theta_0\right)|^{2}> 90\%$ in the practical angular range $-60^{\circ}\leq\theta_0\leq 60^{\circ}$, and $|r_{\mathrm{tri},z_{\mathrm{PMC}}}\left(\theta_0\right)|^{2}> 80\%$ at a wider range, $-80^{\circ}\leq\theta_0\leq 80^{\circ}$, ensues. The reflection phase at the effective PMC plane $z=z_{\mathrm{PMC}}$ [bottom subplot in Fig.\ \ref{Fig:PMC_Results}(b)] remains appreciably stable near the desired goal value $\angle \bar{r}_{\mathrm{tri},z_{\mathrm{PMC}}}\left(\theta_0\right)\equiv 0$, that is, slightly varies between $\angle \bar{r}_{\mathrm{tri},z_{\mathrm{PMC}}}\left(0\right)\approx -6.724^{\circ}\approx -0.117$ rad and $\angle \bar{r}_{\mathrm{tri},z_{\mathrm{PMC}}}\left(\pm 85^{\circ}\right)\approx -0.526^{\circ}\approx0.0087$ rad. This full-wave reflection-phase performance becomes even closer to the ideal PMC response, namely, $|\angle \bar{r}_{\mathrm{tri},z_{\mathrm{PMC}}}\left(\theta_0\right)|<0.13^{\circ}\approx0.0023$ rad for $|\theta_0|\leq 85^{\circ}$, when considering a very-slightly fine-tuned location of the PMC plane, $z_{\mathrm{PMC}}=0.0876$ mm $\approx 3.449$ mil $\approx 0.0058\lambda_0$. Overall, the excellent phase stability observed in our results is of great importance in antenna grounding applications, as it preserves constructive in-phase interference along a wide portion of the spatial spectrum. We have therefore found respectable full-wave performance and agreement with theory, which validate our results for this subsection.

Before concluding, we note that the resonant behavior seen in Fig.\ \ref{Fig:PMC_Results} for the PMC structure elucidates yet another fundamental property of bianisotropy from a different perspective. In view of ($\ref{Eq:scatTE}$) and ($\ref{Eq:Coeff}$), a PMC plane lying at $z=0$ is equivalent to a purely bianisotropic MS of $\widetilde{\chi}_{\mathrm{em}}^{yx}=-\widetilde{\chi}_{\mathrm{me}}^{xy}=-2j$ and $\widetilde{\chi}_{\mathrm{ee}}^{yy}=\widetilde{\chi}_{\mathrm{mm}}^{xx}=\widetilde{\chi}_{\mathrm{mm}}^{zz}=0$ located therein \cite{Tiukuvaara2022}. Consequently, the PMC MS in Fig.\ \ref{Fig:PMC_Config} is equivalent to such a purely bianisotropic MS placed at $z=z_{\mathrm{PMC}}$. 

The properties of such configurations devoid of electric and magnetic susceptibilities ($\bar{\bar{\chi}}_{\mathrm{ee}}=\bar{\bar{\chi}}_{\mathrm{mm}}=0$) have been studied in \cite{Albooyeh2016} (see also Chapter 7.1 in \cite{Achouri2021}). Specifically, it has been found \cite{Albooyeh2016} that passive and lossless purely bianisotropic scatterers cannot exist, i.e., $\bar{\bar{\chi}}_{\mathrm{ee}}=\bar{\bar{\chi}}_{\mathrm{mm}}=0\Rightarrow \bar{\bar{\chi}}_{\mathrm{em}}=\bar{\bar{\chi}}_{\mathrm{me}}=0$; thereby, parasitic electric or magnetic responses are inevitable when bianisotropy is involved; alternatively, loss must be considered as an inherent part of the design.

This insight takes form in our approximations in (\ref{Eq:rPMC_Reduced}) and (\ref{Eq:rPMCexp}): in order to achieve accurate PMC operation at all angles the Pad\'{e} approximant must become exact, namely, $\xi_{\mathrm{mid}}\to\infty$, i.e., $\widetilde{Y}_{\mathrm{mid}}\to\infty$ [see (\ref{Eq:xi})]. Following (\ref{Eq:Equivalence}), this is equivalent to $\widetilde{\chi}_{\mathrm{mm}}^{zz}=0$, which leads, via the grazing-angle HC (\ref{Eq:PMCGrazHC}) to $\xi_{\mathrm{bot}}=0$ and $\widetilde{\chi}_{\mathrm{ee}}^{yy}=0$. In this situation, the \emph{middle} layer is, in fact, a short-circuit, such that the domain above it ($z>0$) is electromagnetically isolated from the domain below ($z<0$) and, hence, altogether irrelevant to the scattering behavior at $z<0$ (and so does $\widetilde{\chi}_{\mathrm{mm}}^{xx}$). In this scenario, the MS effectively reduces to a double layered structure of half the original thickness, where the original middle layer plays the role of the top layer in the new configuration. However, this consequence contradicts the foundations of our framework: the role of the middle layer is distinct from the top one; it \emph{must} interact with the constituents above (and below) it to provide its unique form of contribution to the overall response.

The only option to avoid this futile contradiction is to relinquish the demand for ideal PMC operation by accepting an approximated response via large [e.g., (\ref{Eq:PMC_NoBot})], yet finite, admittance values for the middle layer. In this case, the middle layer becomes highly, but not totally, reflective, such that an effective overcoupled cavity forms between the middle and top layer, in which multiple roundtrip echos take place while leaking as backscattering, especially at the near-grazing angle, in agreement with our results in Fig.\ \ref{Fig:PMC_Results}. This high-finesse phenomenon is inherently resonant and, therefore, narrow-banded and lossy. To conclude this section, we have meticulously demonstrated and validated the viability, practicality, and accuracy of the above concepts.

\section{Conclusion}
To conclude, we have established a powerful and practical scheme to implement thin homogenized MSs of tangential and normal responses \emph{for all angles} via multilyered PCB stacks of tangentially polarizable layers. Particularly, we have unearthed a pivotal analytical connection between these realistic PCB-compatible MSs of cascaded admittance sheets and their equivalent homogenized meta-atom-level counterparts. Unraveling the contribution of each MS layer to each of the effective surface susceptibilities, we have rigorously shown that the normal components are an integral part of such nonlocal structures, which had not been considered in past related work. 

Furthermore, we have demonstrated the profound utility of this formalism, which supports a wide variety of angular filtering functionalities on demand. Specifically, we have harnessed it to meticulously design two exemplary functionalities: an omnidirectionally transparent GHC MS radome and an artificial PMC plane of all-angle in-phase reflection. We have validated the resultant structures, and, thereby, the overall theory herein, by observing excellent performance in simulation and experiment. Along the progress, we have also encountered and discussed several known fundamental and engineering insights congruently captured by our framework herein as well.

Overall, this paper heralds a sound methodology to unite two distinct physical agents of nonlocality, which may surrogate one another. Furthermore, it is quite reasonable to assume that our findings will inspire other similar forms of equivalence between other nonlocal phenomena; these may even reach the realm of optical wavelengths, which are currently engineered by other structures of different natures and principles (e.g., photonic crystals). Hence, it is expected to perfect contemporary nonlocal MS devices such as spatial filters, optical analog computers, and spaceplates, by means of extra-carefully tailored inclusions based on comprehensive foundations.

\appendices

\section*{Acknowledgment}
The authors acknowledge D.\ Dikarov of the Communication Laboratory of the Electrical and Computer Engineering Faculty at the Technion for his kind administrative assistance related to fabricating the specimen measured in this paper.

\section{Transmission-Line Derivation of the Metasurface-Level Scattering Coefficients}
\label{App:TLderivation}
To obtain the scattering coefficients of (\ref{Eq:RTcascade})--(\ref{Eq:g}), we analyze the equivalent TL circuit in Fig.\ \ref{Fig:Theoretical_Config}(c). We first find the equivalent load admittance $\widetilde{Y}_{\mathrm{L},d}$ seen at $z=d$ as simple parallel connection of the shunt $\widetilde{Y}_{\mathrm{top}}$ load and the matched $z>d$ free-space TL, $\widetilde{Y}_{0}=\widetilde{k}_{z,0}$, i.e.,
\begin{equation}
\label{Eq:Y_L,d}
    \widetilde{Y}_{\mathrm{L},d}=\widetilde{Y}_0+\widetilde{Y}_{\mathrm{top}}.
\end{equation}
Next, we apply a standard formula (see, for example, Equation (2.67) in \cite{Pozar2012}) to transform this load across a $d$-long TL of $\widetilde{k}_{z,\mathrm{d}}$ propagation factor and $\widetilde{Y}_{\mathrm{d}}=\widetilde{k}_{z,\mathrm{d}}$ characteristic admittance and find the equivalent input admittance at $z=0$,
\begin{equation}
\label{Eq:Y_in,0}
    \widetilde{Y}_{\mathrm{in},0}=\frac{\widetilde{Y}_{\mathrm{L},d}+j\widetilde{Y}_{\mathrm{d}}\tan\left(k_{z,\mathrm{d}}d\right)}{1+j\widetilde{Y}_{\mathrm{L},d}\widetilde{Y}_{\mathrm{d}}^{-1}\tan\left(k_{z,\mathrm{d}}d\right)}=\frac{h_{\mathrm{top}}^{+}\!\left(\widetilde{k}_{z,0}\right)}{h_{\mathrm{top}}^{-}\!\left(\widetilde{k}_{z,0}\right)},
\end{equation}
where we define the auxiliary terms $h_{\mathrm{top}}^{\pm}\!\left(\widetilde{k}_{z,0}\right)$ as in (\ref{Eq:AuxTerms}); these terms appear after substitution of (\ref{Eq:Y_L,d}) in (\ref{Eq:Y_in,0}). Next, we add this input admittance to the shunt middle load $\widetilde{Y}_{\mathrm{mid}}$ to obtain the equivalent load admittance at $z=0$,
\begin{equation}
\label{Eq:Y_L,0}
    \widetilde{Y}_{\mathrm{L},0}=\widetilde{Y}_{\mathrm{in},0}+\widetilde{Y}_{\mathrm{mid}}.
\end{equation}
Subsequently, we transform this load admittance through another dielectric TL of length $d$ and find the input admittance at $z=-d$,
\begin{equation}
\label{Eq:Y_in,-d}
        \widetilde{Y}_{\mathrm{in},-d}=\frac{\widetilde{Y}_{\mathrm{L},0}+j\widetilde{Y}_{\mathrm{d}}\tan\left(k_{z,\mathrm{d}}d\right)}{1+j\widetilde{Y}_{\mathrm{L},0}\widetilde{Y}_{\mathrm{d}}^{-1}\tan\left(k_{z,\mathrm{d}}d\right)}=\frac{h_{\mathrm{top}}^{+}+h_{\mathrm{top}}^{-}l_{\mathrm{mid}}^{+}}{jg^{-}h_{\mathrm{top}}^{+}+h_{\mathrm{top}}^{-}l_{\mathrm{mid}}^{-}},
\end{equation}
where $l_{\mathrm{mid}}^{\pm}\!\left(\widetilde{k}_{z,0}\right)$ and $g^{-}\!\!\left(\widetilde{k}_{z,0}\right)$ are defined in (\ref{Eq:AuxTerms}) and (\ref{Eq:g}) (they simply appear after several algebraic manipulations; we omit the $\widetilde{k}_{z,0}$-dependence notation for conciseness). Next, we find the total load admittance at $z=-d$ via
\begin{equation}
\label{Eq:Y_L,-d}
        \widetilde{Y}_{\mathrm{L},-d}=\widetilde{Y}_{\mathrm{in},-d}+\widetilde{Y}_{\mathrm{bot}}=\frac{h_{\mathrm{top}}^{+}l_{\mathrm{bot}}^{-}+h_{\mathrm{top}}^{-}\!\left(l_{\mathrm{mid}}^{+}+\widetilde{Y}_{\mathrm{bot}}l_{\mathrm{mid}}^{-}\right)}{jg^{-}h_{\mathrm{top}}^{+}+h_{\mathrm{top}}^{-}l_{\mathrm{mid}}^{-}},
\end{equation}
where $l_{\mathrm{bot}}^{-}\!\left(\widetilde{k}_{z,0}\right)$ is also defined in (\ref{Eq:AuxTerms}). Finally, the reflection coefficient is obtained, as in (\ref{Eq:RTcascade}), via
\begin{equation}
\label{Eq:rTriAppendix}
    r_{\mathrm{tri}}\left(\theta_0\right)=\frac{\widetilde{Y}_{0}-\widetilde{Y}_{\mathrm{L},-d}}{\widetilde{Y}_{0}+\widetilde{Y}_{\mathrm{L},-d}}.
\end{equation}

With this reflection coefficient at hand, we may further find the transmission coefficient. First, we note that the total electric field (equivalent voltage) at $z=-d$ is
\begin{equation}
\label{Eq:VoltageAt-d}
    E_{y}\left(x=0,y=0,z=-d\right)=E_0\left[1+r_{\mathrm{tri}}\left(\theta_0\right)\right]
\end{equation}
(superposition of the incident and reflected waves). This voltage serves as the input voltage for the $-d<z<0$ TL terminated by an effective load of $\widetilde{Y}_{\mathrm{L},0}$ admittance. Hence, by following a simple TL exercise, we may find the resultant electric field at $z=0$ (the load point),
\begin{equation}
\label{Eq:VoltageAt0}
    E_{y}\left(x=0,y=0,z=0\right)=\frac{E_{y}\left(z=-d\right)\sec\left(k_{z,\mathrm{d}}d\right)}{1+\widetilde{Y}_{\mathrm{L},0}\widetilde{Y}_{\mathrm{d}}^{-1}\tan\left(k_{z,\mathrm{d}}d\right)}.
\end{equation}
This voltage (electric field), in turn, serves as the input voltage to the $0<z<d$ TL section. Therefore, identically to (\ref{Eq:VoltageAt0}), we may find the transmitted electric field at $z=d$ via
\begin{equation}
\label{Eq:VoltageAtd}
    E_{y}\left(x=0,y=0,z=d\right)=\frac{E_{y}\left(z=0\right)\sec\left(k_{z,\mathrm{d}}d\right)}{1+\widetilde{Y}_{\mathrm{L},d}\widetilde{Y}_{\mathrm{d}}^{-1}\tan\left(k_{z,\mathrm{d}}d\right)}.
\end{equation}
The transmission coefficient,
\begin{equation}
\label{Eq:Ttri}
    {t_{\mathrm{tri}}\left(\theta_0\right)=}E_{y}\left(x=0,y=0,z=d\right)/E_0,
\end{equation}
in (\ref{Eq:RTcascade}) is then found through consecutive substitutions of (\ref{Eq:VoltageAt-d}) in (\ref{Eq:VoltageAt0}) and (\ref{Eq:VoltageAt0}) in (\ref{Eq:VoltageAtd}) and the definitions of (\ref{Eq:AuxTerms}) and (\ref{Eq:g}).

\section{Extraction of the Admittance Associated With the Meander-Line Layers in Section \ref{Subsubsec:GHC Full-wave}}
\label{App:Characterization}
This appendix accounts for the full-wave method used in Sec.\ \ref{Subsubsec:GHC Full-wave} for obtaining the LUTs  $\mathrm{Im}[\widetilde{Y}(W_L)]$. Starting from the top and bottom layers, we substitute $\widetilde{Y}_{\mathrm{top}}=\widetilde{Y}_{\mathrm{bot}}$ in (\ref{Eq:RTcascade}) and (\ref{Eq:AuxTerms}); following several algebraic manipulations, we find that the associated common susceptance value can be extracted from the simulated results, $r(0;W_{L,\mathrm{top}})$ and $t(0;W_{L,\mathrm{top}})$, via
\begin{equation}
\label{Eq:CharYtop}
    \mathrm{Im}\left[\widetilde{Y}_{\mathrm{top}}(W_{L,\mathrm{top}})\right]=\mathrm{Im}\left[\frac{j}{g^{-}(1)}+\frac{\alpha^{-}+1+g^{-}(1)g^{+}(1)}{\alpha^{+}-1-g^{-}(1)g^{+}(1)}\right],
\end{equation}
where $g^{\pm}(1)$ can be calculated from (\ref{Eq:g}) by substituting $\widetilde{k}_{z,0}=\cos(0)=1$ (normal incidence) and the selected values of substrate parameters in Fig.\ \ref{Fig:GHMS_Config}, and $\alpha^{\pm}$ are defined via
\begin{equation}
\label{Eq:alpha}
    \alpha^{\pm}\left(0;W_{L,\mathrm{top}}\right)=\frac{1\pm r\left(0;W_{L,\mathrm{top}}\right)}{t\left(0;W_{L,\mathrm{top}}\right)}\sec^{2}\left(k_0d\sqrt{\epsilon_{\mathrm{r}}}\right)
\end{equation}
(also using the chosen substrate parameters in Fig.\ \ref{Fig:GHMS_Config}). In the simulations for $r(0;W_{L,\mathrm{top}})$ and $t(0;W_{L,\mathrm{top}})$, we remove the middle layer (yet retain the bonding layer); however, note that (\ref{Eq:CharYtop}) and (\ref{Eq:alpha}) are generally derived from (\ref{Eq:RTcascade}) and (\ref{Eq:AuxTerms}) without any specific assumption regarding the value of $\widetilde{Y}_{\mathrm{mid}}$.

With the top meander width $W_{L,\mathrm{top}}$ fixed according to (\ref{Eq:CharYtop}) and (\ref{Eq:alpha}) and the new full-wave scattering coefficients, $r(0;W_{L,\mathrm{mid}})$ and $t(0;W_{L,\mathrm{mid}})$ simulated in CST with all the layers present (see Sec.\ \ref{Subsubsec:GHC Full-wave}), we may algebraically manipulate (\ref{Eq:RTcascade}) and (\ref{Eq:AuxTerms}) in a different way to extract the admittance associated with the middle layer,
\begin{equation}
\label{Eq:CharYmid}
    \begin{aligned}
        &\mathrm{Im}\left[\widetilde{Y}_{\mathrm{mid}}\left(W_{L,\mathrm{mid}}\right)\right]\\
        &=\mathrm{Im}\left\{ \frac{j}{g^{-}(1)}\left[2-\frac{1+g^{-}(1)g^{+}(1)+\alpha^{+}}{1+jg^{-}(1)+jg^{-}(1)\tilde{Y}_{\mathrm{top}}}\right]\right\},
    \end{aligned}
\end{equation}
where $g^{\pm}(1)$ is calculated as in (\ref{Eq:CharYtop}); $\alpha^{\pm}$ are now calculated by substituting the new full-wave scattering coefficients, $r(0;W_{L,\mathrm{mid}})$ and $t(0;W_{L,\mathrm{mid}})$ in (\ref{Eq:alpha}); and $\widetilde{Y}_{\mathrm{top}}$ is calculated from the expression provided in (\ref{Eq:CharYtop}) with the current values of $\alpha^{\pm}$ (without taking the imaginary part).





\end{document}